\documentclass{jfm}
\usepackage{graphicx}
\usepackage{color,soul}
\usepackage{tikz}
\usepackage{soul}
\usepackage{amsmath,amssymb}
\usepackage{mathrsfs}
\newcommand{\Reg}{\mbox{\textit{Re}}_g}
\newcommand{\Sc}{\mbox{\textit{Sc}}}
\newcommand{\Gr}{\mbox{\textit{Gr}}}
\newcommand{\Fr}{\mbox{\textit{Fr}}}

\renewcommand{\vec}[1]{\boldsymbol{#1}}

\newcommand{\CB}[1]{\textcolor{black}{#1}}

\newcommand{\ti}[1]{\tilde{#1}}

\shorttitle{Transitions in stratified shear flows}
\shortauthor{M. Duran-Matute, S.J. Kaptein, and H.J.H. Clercx}

\title{Regime transitions in stratified shear flows: the link between horizontal and inclined ducts}

\author{M. Duran-Matute\aff{1}
 \corresp{\email{m.duran.matute@tue.nl}}, S.J. Kaptein\aff{1}, \and H.J.H. Clercx\aff{1}}

\affiliation{\aff{1} Fluid Dynamics Laboratory and J.M. Burgers Centre, Department of Applied Physics, Eindhoven University of Technology, PO Box 513, 5600 MB, Eindhoven, The Netherlands}

\begin{document}

\maketitle

\begin{abstract}
\CB{We present the analytical solution for the two-dimensional velocity and density fields within an approximation for laminar stratified inclined duct (SID) flows where diffusion dominates over inertia in the along-channel momentum equation but it is negligible in the density} transport equation. We refer to this approximation as the hydrostatic/gravitational/viscous in momentum and advective in density (HGV-A) approximation due to the leading balances in the governing equations. The analytical solution is valid for laminar flows \CB{in a two-layer configuration} in the limit of long ducts. Under such conditions, \CB{the non-dimensional volume flux is given by the Froude number $Fr^* =\Reg/(A\,K)$} with $\Reg$ the gravitational Reynolds number, \CB{$A$} the aspect ratio of the duct, and $K$ a geometrical parameter that depends on the tilt of the duct \CB{and is obtained from the analytical solution}. The analytical solution in the HGV-A approximation is validated against results from laboratory experiments, and allows us to gain new insight into the dynamics and properties of SID flows. Most importantly, constant values of $Fr^*$ \CB{describe, in both horizontal and inclined ducts, the transitions between increasingly turbulent flow regimes: from laminar flow, to interfacial waves, to intermittent turbulence and sustained turbulence.}
\end{abstract}


\section{Introduction}
The understanding and improved modelling of stratified turbulent flows is of the utmost importance in several environmental situations. A particular flow configuration that attracts a lot of attention is the stratified shear flow. The interest in this flow is twofold. On the one hand, it is reminiscent of several flows occurring in the environment such as the estuarine gravitational circulation \citep{Geyer2014},  the lock exchange flow \citep{Ottolenghi2016,Hartel2000}, and the exchange flow occurring in ocean straits \citep{Gregg1999}. On the other hand, it presents rich dynamics encompassing the emergence of instabilities \citep{Kaminski2014,Ducimetiere2021EffectsInstabilities}, Holmboe waves \citep{Salehipour2016,Lefauve2018}, and stratified turbulence \citep{Salehipour2020,Smith2021TurbulenceFlows,Lefauve2020,Lefauve2022ExperimentalFractions,Lefauve2022ExperimentalParameterisation}. 

\CB{The pre-eminent experimental setup is the stratified inclined duct (SID) due to a high degree of control to explore different flow regimes and phenomena, and to the possibility of performing detailed measurements \citep{Macagno1961,Meyer2014,Lefauve2019,Lefauve2020,Lefauve2022ExperimentalFractions,Lefauve2022ExperimentalParameterisation}. This setup consists of two large tanks with fluid of different densities that are linked by an inclined, long duct (see figure~\ref{fig:setup}).} In recent years, there has been vast progress in the understanding of the flow in SID experiments due to improved measurement capabilities that allow for simultaneous detailed measurements of the three-dimensional (3D) density and velocity fields \citep{Partridge2019}. A central research line has been the transitions between flow regimes: from laminar to the emergence of interfacial waves, to intermittently turbulent, and to fully turbulent \citep{Macagno1961,Meyer2014,Lefauve2019,Lefauve2020}. Although these different regimes have been observed for 60 years, explaining them over a wide range of parameter values and determining the functional dependence of the regime transition on the governing parameters remains \CB{an unsolved problem}. In fact, one of the unanswered questions is ``How to explain flow regime transitions in horizontal ducts or ducts inclined at a slightly negative angle?'' \citep{Lefauve2019}.

\citet{Lefauve2019} distinguished between two situations: \emph{lazy} and \emph{forced} flows. To explain this distinction, it is necessary to define the internal angle of the duct 
$\alpha = \arctan(H/L)$, where $H$ is the height of the duct and $L$ its length. \emph{Lazy} flows are defined as those occurring when the inclination angle $\theta$ of the duct satisfies $\alpha \gg \theta > -\alpha$, and \emph{forced} flows as those occurring when $\theta>\alpha$. Between lazy and forced flows, a smooth transition occurs. The term \emph{forced} refers to the increased importance of the gravitational forcing due to the duct's tilt. \citet{Meyer2014} and \citet{Lefauve2019} have proposed two different \CB{criteria} for the regime transitions in \emph{forced} flows showing good agreement with experimental data. \CB{However, the criterion proposed by \citet{Meyer2014} is \emph{a priori} not valid for $\theta \leq 0$, and the one proposed by \cite{Lefauve2019} is \emph{a priori} not valid for lazy flows.}

The current paper proposes a new \CB{criterion for the regime transitions in SID experiments that spans both lazy and forced flows (i.e. encompassing slightly negative inclinations, horizontal ducts, and positive inclinations). This criterion is based on a formal perturbation analysis for long ducts and the analytical solution of the resulting simplified set of equations. This work builds upon the recent description by \citet{Kaptein2020} of what they called the \emph{high-advection/low-diffusion} approximation for laminar flows in horizontal ducts. In this approximation, viscous diffusion dominates over inertia in the along-channel momentum equation while diffusion is negligible in the density transport equation. In fact, the flow gets organized into two layers with a sharp interface in between. For consistency with the work by \citet{Lefauve2020}, we will refer here to this approximation as the HGV-A approximation because of the hydrostatic/gravitational/viscous balance in the momentum equation and the dominance of advection in the density transport equation (a detailed derivation is presented in Sec. \ref{sec:scaling}). This approximation is, in general, only possible if the Schmidt number (the ratio between the kinematic viscosity of the fluid and the diffusivity of the scalar responsible for the density differences) is much larger than unity, for example, for salt-stratified flows. \cite{Kaptein2021ReproducingDomains} proposed new curves based on this approximation describing the regime transitions with a generalized Reynolds number. These curves show an overall better agreement with experimental results for salt-stratified flows than previously proposed criteria. However, they are based on an empirical assumption of the slope of the interface as a function of the ratio $\theta/\alpha$, which impedes a thorough physical explanation.}\\
To provide such an explanation and propose transition curves based on physical principles, we derive, in the current paper, a two-dimensional (2D) analytical solution for \CB{the velocity and density fields} in the HGV-A approximation in both horizontal and slightly inclined ducts. This analytical solution provides several new insights into SID flows. In particular, it provides the non-dimensional parameter governing the regime transition for SID flows for horizontal and inclined ducts. \CB{The new understanding of SID flows should allow better targeted experimental campaigns to answer remaining questions due to the clearer view of the parameter space. Furthermore, the tilt angle in relatively small-scale, well-controlled SID experiments can be used to achieve more turbulent flows than in horizontal ducts with the same values of the other governing parameters. This means that understanding the link between horizontal and inclined ducts can help extend the results of SID experiments in an inclined duct in the fully-turbulent regime to large-scale environmental exchange flows, which are mostly horizontal (and turbulent).}

\section{Description of the system and background}
\begin{figure}
    \centering
    \includegraphics[width=0.8\textwidth]{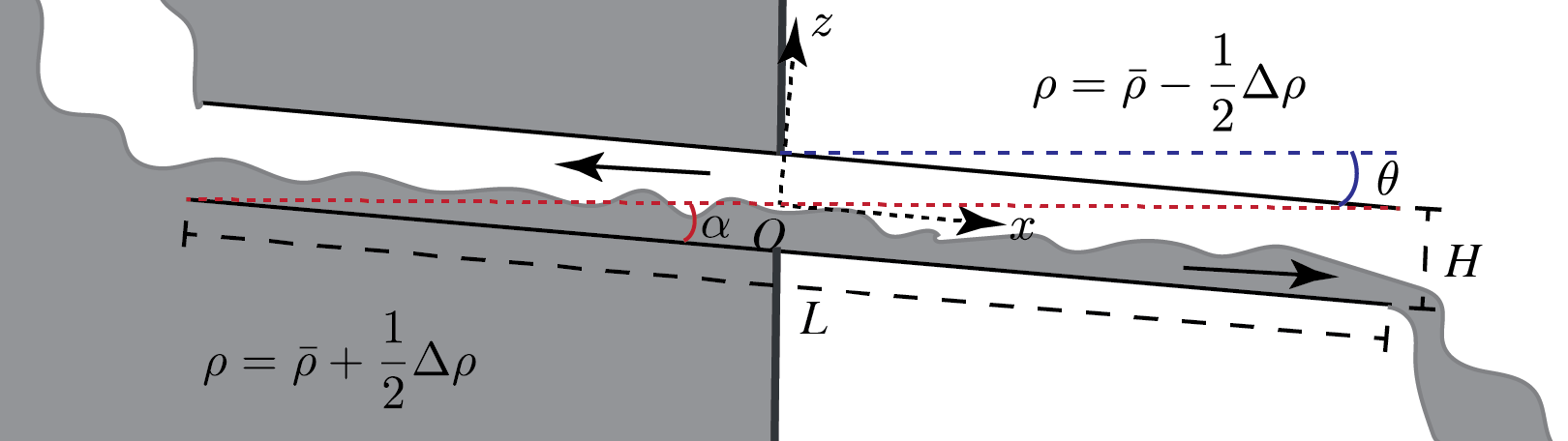}
    \caption{Schematic representation of the side view of a stratified inclined duct (SID) experimental setup. The duct of length $L$ and height $H$ is inclined at an angle $\theta$ with respect to the horizontal. The duct connects two large tanks: one with water with density $\rho=\bar{\rho}+\Delta\rho/2$ and the other with density $\rho=\bar{\rho}-\Delta\rho/2$. The internal angle of the duct is $\alpha=\arctan(H/L)$. The along-duct coordinate is $x$ and the coordinate perpendicular to the bottom and the top of the duct is $z$. The origin $O$ of the coordinate system is located at the center of the duct.}
    \label{fig:setup}
\end{figure}

The SID setup, mentioned earlier and sketched in figure~\ref{fig:setup}, consists of two tanks with fluid at densities $\bar{\rho}\pm \Delta \rho/2$ (due to differences in, for example, salt concentration or temperature), joined by a duct. The duct has length $L$ and height $H$, and it is inclined at an angle $\theta$ with respect to the horizontal. The fluid is considered to have uniform and constant viscosity $\nu$. It is convenient to define the buoyancy velocity scale $U_g \equiv \sqrt{g' H}$, where $g'\equiv g \Delta \rho /\bar{\rho}$ with $g$ the gravitational acceleration. Besides the \CB{inclination} angle of the duct $\theta$, the system can be described by three non-dimensional parameters: the aspect ratio of the duct \CB{$A \equiv \cot \alpha  = L/H$}, the gravitational Reynolds number
\begin{equation}
    \Reg \equiv\frac{H U_g}{2\nu}= \frac{H \sqrt{g' H}}{2\nu},
\end{equation}
and the Schmidt number $Sc \equiv \nu/\kappa$ with $\kappa$ the diffusivity of salt (or heat, in which case the Schmidt number is referred to as the Prandtl number). We consider long ducts \CB{($A \gg 1$) for which $A^{-1}\approx \alpha$.} For ducts with finite width $W$, we must introduce an additional parameter $B\equiv W/H$. 

\citet{Meyer2014} proposed an empirical criterion for the transition between different flow regimes by defining the Grashof number as $\Gr \equiv 2A \Reg^2\sin\theta $ which quantifies the ratio of the buoyancy force to the viscous force. They proposed the critical value $\Gr= 4\times 10^7$ for the transition between the intermittently turbulent and the turbulent regimes showing good agreement with experimental results. 
\citet{Lefauve2019} proposed that the transitions between different regimes for a SID setup with a given $A$-value occur at constant $\theta \Rey_g$-values. \citet{Lefauve2020} \CB{checked the proposed transitions at constant $\theta \Rey_g$-values} against several experimental data sets including those of \citet{Meyer2014}. They remarked particularly good agreement with experiments for \emph{forced} flows ($\theta>\alpha$) when, in addition, $A^{-1} \Reg \lesssim 50 $. However, the comparison was inconclusive for other values of $\theta$ and $A^{-1} \Rey_g$. 

Both previously mentioned criteria [those proposed by \citet{Meyer2014} and \citet{Lefauve2019}] have a crucial shortcoming: they are not valid for $\theta \leq 0$ because the proposed governing parameters ($\Gr$ and $\theta \Reg$) are then equal or smaller than zero. This would mean that for $\theta\leq 0$, the flow does not transition away from laminar, which is in disagreement with experimental results \CB{\citep[see e.g. Figure 4 by][]{Lefauve2020}}. Moreover, it is well known that the Grashof number defined as $\Gr = \Reg^2$ is the governing parameter in the case of a horizontal duct \citep[see e.g.][]{Hartel2000,Hogg2001}. \CB{More precisely, the governing parameter is $\Gr A^{-2} = \Reg^2 A^{-2}$ according to \citet{Hogg2001}.} Hence, the definition of $\Gr$ proposed by \citet{Meyer2014} is inconsistent with the relevant definition for horizontal ducts.  For these reasons, there is still a need to find a physical explanation and a generalized governing parameter \CB{that determines the transitions which is valid for positively inclined, horizontal, and negatively inclined ducts.}

\CB{From the side of horizontal ducts, it is known} that diffusion dominates and $U/U_g\propto \Reg A^{-1}$ (with $U$ denoting the typical magnitude of the velocity) for  $\Reg A^{-1} \ll  (180/ \Sc)^{1/2}$ \CB{\citep{Kaptein2020}.} This approximation is known as the \emph{viscous advective-diffusive (VAD) solution} \citep{Cormack1974,Hogg2001}, the \emph{hydrostatic-viscous balance} \citep{Lefauve2020} or \emph{the diffusion-dominated regime} \citep{Kaptein2020}. In the opposite case of  $\Reg A^{-1}\gg  (180/ \Sc)^{1/2}$, \citet{Kaptein2020} found two distinct approximations that arise depending on $\Sc$\CB{: one for $Sc\approx 1$ and another for $Sc\gg 1$.} 

For $\Sc\approx 1$, the flow tends to the hydraulic limit, in which $U/U_g\propto 1$ \citep{Hogg2001}. This is the theoretical limit for a steady, inviscid, irrotational, hydrostatic flow in which the two layers are of equal thickness all along the duct \citep{Gu2005,Lefauve2020}. \CB{For $\Sc\gg 1$, \citet{Kaptein2020} found that $U/U_g\propto \Reg A^{-1}$, similarly to the VAD solution but with a different proportionality constant. This behavior would correspond to what we refer here as the HGV-A approximation. However, the observed scaling cannot continue indefinitely as the value of $\Reg$ increases, since the hydraulic limit also exists for $Sc\gg1$, and hence, there is the upper-bound $U/U_g=1$. Most probably, \citet{Kaptein2020} did not observe the transition towards the hydraulic limit because they only modeled laminar, steady flows. In fact, the limit of the parameter space that they explored was established by the emergence of waves and instabilities at the interface.} This suggests that the HGV-A approximation should hold for flows with $\Sc\gg1$ between the VAD solution \CB{(which is not hydraullically controlled) and a hydraulically controlled flow. It is this hypothesis that we will endeavour to verify in this paper.}

\section{Analytical description of the HGV-A approximation}\label{sec:model}

\CB{In this section, we focus on the HGV-A approximation and derive its consequences for regime transitions in inclined ducts. Although scaling analysis for SID flow has been previously done by \citet{Lefauve2020}, the HGV-A approximation was not considered. Furthermore, we take a slightly different approach in the non-dimensionalization that allows us to simplify the equations in a mathematically formal way by using an asymptotic analysis of the momentum and density transport equations describing the flow in a long duct ($A\gg 1$) \citep[see][for background theory]{VanDyke1975PerturbationMechanics}. Such an analysis has been proven to be a powerful tool to analyze problems where the geometrical shape of the domain introduces a small parameter \citep[see e.g.][]{Rienstra2001AnalyticalConditions,Duran-Matute2012}, which in this case is $A^{-1}$. In fact, \citet{Cormack1974} used a similar approach to derive the VAD approximation in a closed container.\\
 First, in Section \ref{sec:scaling}, we present the non-dimensional governing equations that are later used, in Section \ref{sec:asymptotic}, to derive the HGV-A approximation using asymptotic analysis for a long duct. In Section \ref{sec:analytic}, we present the analytical solution for the two-dimensional velocity and density fields. Then, we discuss, in Section \ref{sec:implications}, its implications for the regime transitions.}

\subsection{Governing equations}\label{sec:scaling}

We consider the 2D, steady flow in a $(x,z)$ cross-section of an infinitely wide duct, where $x$ is the along-channel coordinate and $z$ the coordinate going from the bottom to the top of the duct. The fluid velocity is $\vec{v}=(u,0,w)$. The flow is described by the continuity and the steady Navier-Stokes equations with the Boussinesq approximation for an incompressible fluid:
\begin{eqnarray}
\frac{\partial u}{\partial x}+\frac{\partial w}{\partial z} & = & 0,\label{eq:cont2d} \\
u\frac{\partial u}{\partial x}+w\frac{\partial u}{\partial z} & = & 
-\frac{1}{\bar{\rho}}\frac{\partial p}{\partial x}+\nu \left( \frac{\partial^{2} u}{\partial x^{2}}+\frac{\partial^{2} u}{\partial z^{2}} \right)+g\frac{\rho'}{\bar{\rho}}\sin \theta, \label{eq:ns2dx} \\
u\frac{\partial w}{\partial x}+w\frac{\partial w}{\partial z} & = & 
-\frac{1}{\bar{\rho}}\frac{\partial p}{\partial z}+\nu \left( \frac{\partial^{2} w}{\partial x^{2}}+\frac{\partial^{2} w}{\partial z^{2}} \right) - g\frac{\rho'}{\bar{\rho}}\cos \theta, \label{eq:ns2dz}
\end{eqnarray}
\CB{where $p$ is the pressure and $\rho'=\rho-\bar{\rho}$ is the variable part of the total density, with 
$\bar{\rho} \gg \rho'$. In the reference frame of the inclined duct, the gravity vector is $\vec{g}=(g \sin\theta, 0, -g \cos \theta)$.} A linear equation of state relates the density $\rho$ to, for example, salt concentration or fluid temperature. In this way, the density $\rho$ is also governed by a steady transport equation:
\begin{equation}
u\frac{\partial \rho}{\partial x}+ w\frac{\partial \rho}{\partial z}
 =	
\kappa \left( \frac{\partial^{2} \rho}{\partial x^{2}} + \frac{\partial^{2} \rho}{\partial z^{2}} \right). \label{eq:trans2D}
\end{equation}
We define the non-dimensional variables denoted by a tilde such that
\begin{equation}\label{eq:nd_var}
    u=U \tilde{u},\: w=\dfrac{U}{A} \ti{w},\: x=\dfrac{L}{2}\ti{x},\: z=\dfrac{H}{2}\ti{z},\: \rho'=\Delta\rho \,\ti{\rho}', \: p=\dfrac{2\bar{\rho} U \nu L}{H^2} \ti{p}.
\end{equation}

\CB{Using the non-dimensional variables defined in \eqref{eq:nd_var}, (\ref{eq:cont2d})--(\ref{eq:trans2D}) can be written as
\begin{eqnarray}
     \frac{\partial \ti{u}}{\partial \ti{x}}+\frac{\partial \ti{w}}{\partial \ti{z}}&=&0,\label{eq:cont_nd}\\
A^{-1} \Rey_g\, \Fr  \left(\ti{u}\frac{\partial \ti{u}}{\partial \ti{x}}+\ti{w}\frac{\partial \ti{u}}{\partial \ti{z}}\right)\label{eq:cont2d_nd} & = & 
-\frac{\partial \ti{p}}{\partial \ti{x}}+A^{-2} \frac{\partial^{2} \ti{u}}{\partial \ti{x}^2}+\frac{\partial^{2} \ti{u}}{\partial \ti{z}^2}+\dfrac{\Rey_g}{2 \Fr}  \ti{\rho}'\sin \theta, \label{eq:ns2dx_nd} \\
 A^{-3} \Rey_g \Fr  \left(\ti{u}\frac{\partial \ti{w}}{\partial \ti{x}}+\ti{w}\frac{\partial \ti{w}}{\partial \ti{z}}\right) & = & 
-\frac{\partial \ti{p}}{\partial \ti{z}}+A^{-4} \frac{\partial^{2} \ti{w}}{\partial \ti{x}^{2}}+A^{-2}\frac{\partial^{2} \ti{w}}{\partial \ti{z}^{2}} - A^{-1}\dfrac{\Rey_g}{2\Fr}  \ti{\rho}' \cos \theta, \label{eq:ns2dz_nd}\\
   \hspace{-0.5cm} A^{-1}\Rey_g\,\Sc\,\Fr  \left(\ti{u}\frac{\partial \ti{\rho}'}{\partial \ti{x}}+ \ti{w}\frac{\partial \ti{\rho}'}{\partial \ti{z}}\right) & = &
 A^{-2}\frac{\partial^{2} \ti{\rho}'}{\partial \ti{x}^{2}} + \frac{\partial^{2} \ti{\rho}'}{\partial \ti{z}^{2}}, \label{eq:trans2D_nd}
\end{eqnarray}
where the Froude number is defined as 
\begin{equation}\label{eq:Fr}
    \Fr\equiv U/U_g.
\end{equation}
Here, we define the typical velocity scale $U\equiv 2Q/H$, where  
\begin{equation}
    Q\equiv \dfrac{1}{2}\int_{-H/2}^{H/2}|u|dz
\end{equation}
is the dimensionful volume flow rate through the duct in one direction such that $U$ is the average velocity in each direction. We do not set, \emph{a priori}, $U_g$ as the typical velocity scale \citep[as done by e.g. ][]{Cormack1974} because we do not know if $U$ scales with $U_g$, and it is imperative to use a representative velocity scale to fully profit from the asymptotic analysis. In fact, determining the relationship between $U$ (or $Q$) and $U_g$ is a crucial step in deriving the analytical solution in the HGV-A approximation.} \CB{Hence, the Froude number represents the non-dimensional volume flow rate through the duct, and it is related to the function $f_{\Delta U}$ discussed by \citet{Lefauve2020} for which they used the peak-to-peak velocity as the typical velocity scale.}

\subsection{Asymptotic analysis for long ducts}\label{sec:asymptotic}

\CB{We now make use of the fact that $A^{-1}\ll 1$ to describe the main characteristics and balances in the HGV-A approximation using an asymptotic analysis. We introduce the symbol $O$ to denote the mathematical order of a function and $O_s$ to denote the \emph{sharp} order \citep[see e.g.][for the formal definitions]{Ekhaus1979}. Notice that the symbol $O$ does not denote the physical order of magnitude because no account is kept of constants of proportionality \citep{VanDyke1975PerturbationMechanics}. Our approach differs in three ways from that of \citet{Cormack1974}. First, as already mentioned, we do not consider \emph{a priori} $U_g$ as a representative velocity scale. Second, we simplify the problem by considering only the lowest order terms in the expansion. Third, we complicate the problem by not assuming that the interface is parallel to the bottom and top of the duct.}  

\CB{The starting point of the asymptotic analysis is to write the velocity, the pressure and the variable part of the density as asymptotic expansions on $A^{-1}$, such that 
\begin{equation}\label{eq:expansion}
    \tilde{u}=\sum_{n=0}^\infty A^{-n}\tilde{u}_n, \quad
    \tilde{w}=\sum_{n=0}^\infty A^{-n}\tilde{w}_n, \quad
    \tilde{p}=\sum_{n=0}^\infty A^{-n}\tilde{p}_n, \quad
    \tilde{\rho}'=\sum_{n=0}^\infty A^{-n}\tilde{\rho}'_n.
\end{equation}
As a second step, it is necessary to express the response parameters, in this case, $Fr$ as a function of $A^{-1}$. We write
\begin{equation}\label{eq:param_A}
\Fr = C A^{-l},
\end{equation}
where $C$ is a finite constant such that $\Fr =O_s(A^{-l})$ with $l \in N$. We now substitute \eqref{eq:expansion} and \eqref{eq:param_A} into \eqref{eq:cont2d_nd}--\eqref{eq:trans2D_nd}, and extract the equations for the lowest (zeroth order) terms.}

\CB{We start the analysis with the $x$ and $z$ components of the Navier-Stokes equations, \eqref{eq:ns2dx_nd} and \eqref{eq:ns2dz_nd}, respectively. Considering $A^{-1}\ll1$ immediately yields, as expected for long ducts, that horizontal viscous diffusion (terms with $\partial^2/\partial x^2$) can be neglected from these equations when compared to vertical viscous diffusion (terms with $\partial^2/\partial z^2$) because of the additional $A^{-2}$ factor. The lowest order term of vertical viscous diffusion in \eqref{eq:ns2dz_nd} is $O(A^{-2})$ can also be immediately neglected with respect to lowest order terms of the pressure gradient and gravity forces which are $O(A^{0})$.} 

\CB{Different balances can be obtained by considering different values of $l$ in \eqref{eq:param_A}. However, we are here interested only in the laminar flow for which the non-linear terms in \eqref{eq:ns2dx_nd} and \eqref{eq:ns2dz_nd} can be neglected. This means that, formally, we consider here only the case where $\Rey_g \Fr A^{-1}=O(A^{-1})$. Equation \eqref{eq:ns2dz_nd} yields then, at lowest order, the hydrostatic balance:
\begin{equation}\label{eq:ns2dz_nd2}
    \frac{\partial \ti{p}_0}{\partial \ti{z}} = -\dfrac{K}{2} \ti{\rho}'_0\cos \theta
\end{equation}
with
\begin{equation}\label{eq:K}
    K \equiv \dfrac{\Reg}{A\,\Fr}.
\end{equation}
However, the hydrostatic balance \eqref{eq:ns2dz_nd2} can only hold if $K\cos\theta$ is a finite constant (formally, if $K\cos\theta=O_s(A^0)$),
so that the terms on both sides are of the same order. Furthermore, $\theta\ll1$ so $\cos \theta \approx 1$, and thus, $K$ must be a finite constant ($0<K<\infty$) in the limit $A^{-1}\to 0$ (formally, $K=O_s(A^0$)).}

\CB{We now turn our attention to \eqref{eq:ns2dx_nd}, where the remaining terms yield 
\begin{equation}\label{eq:ns2dx_nd2}
   \frac{\partial^{2} \ti{u}_0}{\partial \ti{z}^{2}}=
 \frac{\partial \ti{p}_0}{\partial \ti{x}}-\dfrac{K}{2} \ti{\rho}'_0 A \sin \theta.
\end{equation}
Here, we can clearly see that the hydrostatic pressure and gravity are balanced by viscous momentum diffusion. This is the balance that gives rise to the HGV acronym. Here, the condition that $\theta \ll1$ is also important so that $A \sin \theta$ remains finite when $A\gg1$. Notice that taking $\sin \theta =0$ yields the governing equation for a horizontal duct or what \citet{Lefauve2020} call the hydrostatic/viscous (HV) balance.}
Finally, the $z$-component of the velocity, $\ti{w}$, is given by the continuity equation \CB{
\begin{equation}\label{eq:cont_nd2}
    \frac{\partial \ti{u}_0}{\partial \ti{x}}+ \frac{\partial \ti{w}_0}{\partial \ti{z}}=0.
\end{equation}}

\CB{The simplified equations \eqref{eq:ns2dz_nd2}, \eqref{eq:ns2dx_nd2} and \eqref{eq:cont_nd2} were already introduced by \citet{Macagno1961} as the governing equations for SID flows. However, the perturbation analysis yields the additional important result that $K=O_s(A^0)$. This means that $Fr=O_s(A^{-1})$, i.e. that $\Fr=U/U_g = \Rey_g /( A\,K)$ \citep[in agreement with the numerical results of ][]{Kaptein2020}, with $K^{-1}$ a proportionality constant in the limit of $A^{-1}\to 0$. We can see from \eqref{eq:ns2dx_nd2} that the balance, and hence, the value of $K$ still depends on the value of $A \sin \theta$ which remains a finite parameter in the simplified set of equations. In short, the perturbation analysis so far tells us that, to find the dimensionless volumetric flow rate $Fr= \Rey_g /(K\, A)$, we must find the value of $K$ as a function of $A \sin\theta$ since the values of $Re_g$ and $A$ are known.}

\CB{We now continue our analysis with the density transport equation \eqref{eq:trans2D_nd}, which yields at lowest order:
\begin{equation}\label{eq:trans2D_nd2}
       \frac{\partial^{2} \ti{\rho}'_0}{\partial \ti{z}^{2}} = 0,
\end{equation}
meaning that vertical diffusion of density is equal to zero.}

In short, the HGV-A approximation is characterized by a hydrostatic balance in the vertical, while in the horizontal, the flow is driven by a pressure gradient force, \CB{and by gravity in the along-duct direction if $\theta \neq 0$.} Finally, we have the hallmark of the HGV-A approximation: \CB{diffusion dominates over inertia in the along-channel momentum equation but is equal to zero in the density transport equation}. \CB{For this combination to be possible and this approximation to be relevant, it is needed that $Sc$ is large \citep[as observed by][]{Kaptein2020} which is the case, for example, for water flows where the density difference is caused by a difference in salt concentration ($\Sc\approx 700$) and not by a difference in temperature ($Sc\approx 7$).}\\

\subsection{Analytical solution in the HGV-A approximation}
\label{sec:analytic}

\CB{We consider now the governing equations at lowest order \eqref{eq:ns2dz_nd2} and \eqref{eq:ns2dx_nd2}--\eqref{eq:trans2D_nd2}, referred to as the HGV-A approximation.
For simplicity in the notation, we drop the tildes. The goal now is to find an analytical solution for this set of equations. We split the procedure in two steps. First, we determine the expressions for the along-channel velocity $u_0$ and the density $\rho'_0$. Second, we determine the value of $Fr$ which, as mentioned earlier, is equivalent to finding the value of $K$ as a function of $A \sin \theta$.}


\subsubsection{Determining the along-channel velocity and the density}

To find an expression for the along-channel velocity $u_0$ and the density $\rho'_0$ in the HGV-A approximation, we begin with the transport equation \eqref{eq:trans2D_nd2}. For the solution to this equation, we propose 
\begin{equation}
\displaystyle \rho_0'(x,z) = \frac{1}{2} - \mathcal{H}(z-\eta(x)),
\label{eq:rhoSteadyState-adv}
\end{equation}
where $\mathcal{H}(z)$ is the Heaviside function defined as: $\mathcal{H}(z>0)=1$, 
$\mathcal{H}(z<0)=0$, and $\mathcal{H}(z=0)=1/2$. The physical meaning of this solution is that the density gets organized in two layers with a sharp interface located at $z=\eta(x)$ with $\eta(0)=0$ due to the definition of the coordinate system. 
\CB{The proposed solution satisfies $\partial^2 \rho'_0/\partial z^2=0$ in each of the layers. However, formally, the solution is not defined at the interface, and it is convenient to approach the problem by treating each layer separately, with the interface as a boundary between them.} We use the index $\zeta=\pm 1$ with $\zeta=+1$ referring to the top layer and $\zeta=-1$ to the bottom layer. Using this notation, $\rho'_{0,\zeta}=-\zeta/2$. \CB{\citet{Macagno1961} also solved the same governing equations for a two layer system, but they considered the simplified problem where the interface is parallel to the top and bottom of the duct, i.e. $\eta(x)=0$.}

We now turn our attention to \eqref{eq:ns2dz_nd2} \CB{to derive the pressure distribution and the horizontal pressure gradient that drives the flow}. Integrating this equation with respect to $z$ yields an expression for the pressure:
\begin{equation}\label{eq:pressure}
    p_\zeta(x,z)=K\dfrac{\zeta}{4}z\cos \theta + \gamma_\zeta(x)
\end{equation}
with $\gamma_\zeta(x)$ integration constants, which are functions of $x$ to be determined using the boundary conditions. At $z=1$, the pressure is an unknown function of $x$, but it is convenient to define, without loss of generality, 
\begin{equation}\label{eq:pressure_top}
  p_{0,+1} (x,z=1)=-K[f(x)- \eta(x) \cos\theta]/4   
\end{equation}
with $f(x)$ an unknown function \CB{giving the $x$-dependence of the barotropic pressure}. Using \eqref{eq:pressure} and \eqref{eq:pressure_top}, one obtains an expression for $\gamma_{+1}(x)$ as a function of $\eta(x)$ and $f(x)$, and applying continuity of pressure at $z=\eta(x)$ yields an expression for $\gamma_{-1}(x)$. The pressure as a function of $x$ and $z$ is then given by
\begin{equation}
    p_{0,\zeta}(x,z)=\zeta\dfrac{K}{4}[\cos\theta(z-\eta(x)-\zeta)+\zeta f(x)]
\end{equation}
so that the horizontal pressure gradient is
\begin{equation}\label{eq:dpdx}
\dfrac{\partial p_{0,\zeta}}{\partial x}=\zeta\dfrac{K}{4}\left( \cos\theta \dfrac{d \eta(x)}{d x}-\zeta \dfrac{d f(x)}{d x}\right),
\end{equation}
\CB{where we can see that it is composed of a baroclinic part due to the sloping interface and a barotropic part given by the gradient of $f(x)$.}

Substituting \eqref{eq:dpdx} into \eqref{eq:ns2dx_nd2} yields
\begin{equation}\label{eq:u}
    \frac{\partial ^2 u_{0,\zeta}}{\partial z^2}=  -\zeta\dfrac{K}{4} F_\zeta(x)
\end{equation}
with
\begin{equation}
    F_\zeta(x)=\cos\theta \dfrac{d \eta(x)}{d x}-\zeta\dfrac{d f(x)}{d x}-A\sin \theta.
\end{equation}
Integrating twice with respect to $z$ and applying the boundary conditions $u_0(z=\pm 1)=u_0(z=\eta(x))=0$ gives
\begin{equation}\label{eq:u1}
    u_{0,\zeta}(x,z) =-\zeta\dfrac{K}{8}F_\zeta(x)(z-\zeta)(z-\eta(x)).
\end{equation}

\CB{Since the barotropic pressure gradient is responsible for an equal volume flow rate in both directions, we use the condition of zero mean flow through the duct:
\begin{equation}
    \int_{-1}^{1} u_0(x,z) dz=0
\end{equation}
to determine $d f(x)/dx$. We do this by integrating $u_{0,-1}(x,z)$ from $z=-1$ to $z=\eta(x)$ and $u_{0,+1}(x,z)$ from $z=\eta(x)$ to $z=1$, yielding}
\begin{equation}
    \dfrac{d f(x)}{d x}= -\dfrac{\eta(x)[3+\eta^2(x)]}{1+3\eta^2(x)}\left(\cos\theta \dfrac{d \eta(x)}{d x}-A\sin \theta \right),
\end{equation}
and $F_\zeta(x)$ can be rewritten as
\begin{equation}\label{eq:Fj}
    F_\zeta(x)= \dfrac{[1+\zeta\eta(x)]^3}{1+3\eta^2(x)}\left(\cos \theta \dfrac{d \eta(x)}{d x}-A\sin \theta\right).
\end{equation}
Replacing this expression for $F_\zeta$ into \eqref{eq:u} \CB{clearly shows the physical origin of the two drivers of the flow: (i) a baroclinic pressure gradient due to a sloping interface between the two layers with homogeneous density, and (ii) gravity due to the tilt of the duct. For a given value of $A$, the relative importance of these two forcing terms varies with the angle $\theta$ and the slope of the interface. For example, for horizontal ducts, the pressure gradient will be the only forcing, while it would be expected that gravity takes over with increasing $\theta$ values (particularly, if $d \eta(x)/dx$ decreases simultaneously). Hence, considering these two forcing terms without neglecting \emph{a priori} any of them is crucial to link our knowledge of horizontal and inclined ducts. The difficulty here is that although the value of $\theta$ is known since it is a control parameter, the value of $d \eta(x)/dx$ is not.}  

Now, the only missing part of the solution is to determine the shape of the interface $\eta(x)$. For this, we consider the flow rate through each of the layers yielding
\begin{equation}
    -\int_{\eta(x)}^{1} u_{0,+1}(x,z) dz=1 =\int_{-1}^{\eta(x)}u_{0,-1}(x,z) dz,
\end{equation}
\CB{where the value of one is due to the way the velocity was made dimensionless.}
In this way, we obtain an autonomous differential equation for $\eta(x)$:\CB{
\begin{equation}\label{eq:detadx}
   \cos\theta \dfrac{d \eta(x)}{d x}=-\dfrac{48 }{K}\dfrac{1+3\eta^2(x)}{[1-\eta^2(x)]^3}+A\sin \theta.
\end{equation}}

Equation \eqref{eq:detadx} can be further used to rewrite $F_\zeta(x)$ in \eqref{eq:Fj} yielding
\begin{equation}\label{eq:Fj2}
    F_\zeta(x)= -\dfrac{48 }{K}\dfrac{1}{[1-\zeta\eta(x)]^3},
\end{equation}
such that \eqref{eq:u} is now written as
\CB{\begin{equation}\label{eq:HGV-A_u}
    \frac{\partial ^2 u_{0,\zeta}}{\partial z^2}=  -\dfrac{12\zeta}{[1-\zeta\eta(x)]^3}.
\end{equation}}
Finally, the along-duct velocity given by \eqref{eq:u1} can be written as 
\begin{equation}\label{eq:u_HGV-A_F}
    u_{0,\zeta}(x,z) =\dfrac{6  \zeta}{[1-\zeta\eta(x)]^3}(z-\zeta)(z-\eta(x)),
\end{equation}
\CB{where we can notice that the velocity profile at $x=0$,
\begin{equation}
   u_{0,\zeta}(0,z) =6  \zeta (z-\zeta)z,
\end{equation}
\CB{has always the same shape consisting of two parabolas. To reach the complete solution, it is still necessary to determine the position of the interface, which is equivalent to finding the value of $K$ as a function of $A \sin \theta $, which can be done by solving \eqref{eq:detadx} for $\eta(x)$. }}

\subsubsection{Determining $K$}\label{sec:K}

\CB{Since the analytical solution for horizontal ducts is tractable}, we focus first on this case for which \eqref{eq:detadx} simplifies to
\begin{equation}\label{eq:detadx_h}
    \dfrac{d \eta(x)}{d x}=-\dfrac{48 }{K }\dfrac{1+3\eta(x)^2}{[1-\eta(x)^2]^3}, 
\end{equation}
yielding 
\begin{equation}\label{eq:int_hor}
    x=-\dfrac{K}{19440}[320\sqrt{3} \arctan (\sqrt{3} \eta(x))-555 \eta(x)+150 \eta(x)^3 - 27 \eta(x)^5].
\end{equation}

\begin{figure}
    \centering
    \includegraphics[width=0.95\textwidth]{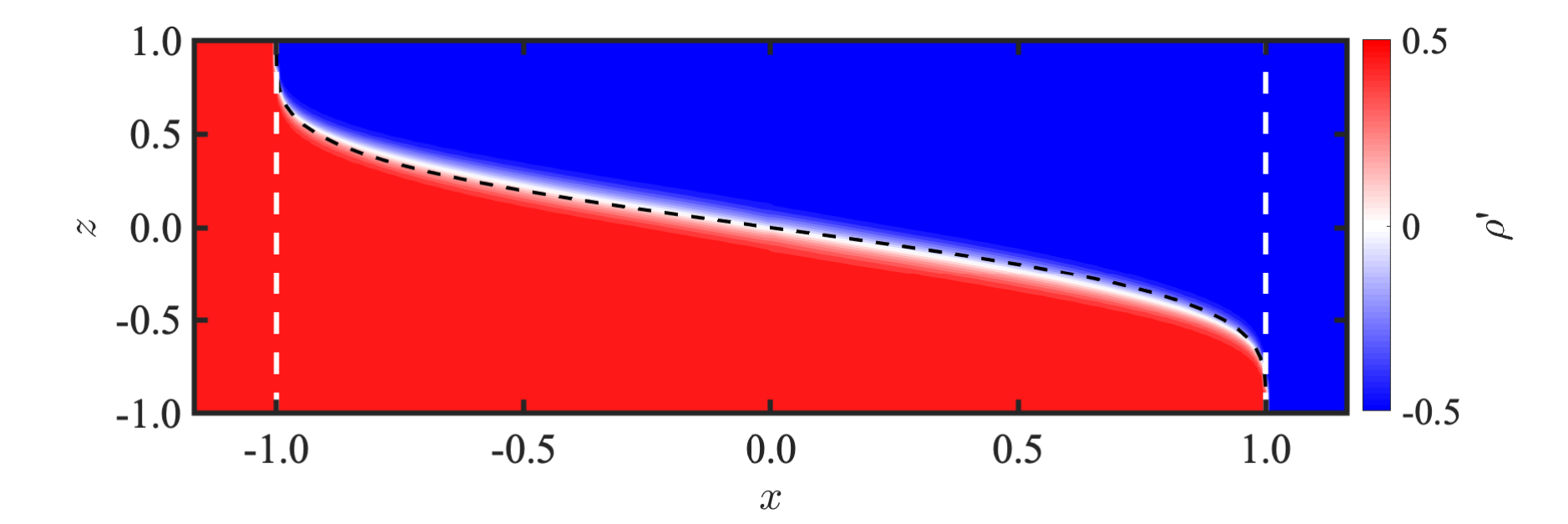}
    \caption{Density field from the numerical simulation by \citet{Kaptein2020} for \CB{$\Reg=500$}, $A=60$, $\Sc=300$, and $\theta=0$. The white dashed lines represent the limits of the duct ($x=\pm 1$). The black dashed line represents the interface given by \eqref{eq:int_hor} with $K=131$.}
    \label{fig:rho_sim}
\end{figure}

The value of $K$ is obtained by imposing boundary conditions. \CB{Finding the slope of the interface from an autonomous equation---similar to \eqref{eq:detadx}---was previously done for horizontal ducts by \cite{Gu2005} and for inclined ducts by \cite{Lefauve2020} who derived the equation using internal hydraulics. \citet{Gu2005} found in fact a similar expression to \eqref{eq:int_hor}}, but the constant to be determined from the boundary conditions was the composite Froude number given a certain magnitude of the frictional effects. When using internal hydraulics, the boundary condition is proposed using maximum exchange flow theory \citep{Armi1986MaximalFlow}. This theory states that the maximum flow rate possible is such that the flow becomes critical at the edges of the channel (i.e. that the composite Froude number becomes unity at $x=\pm1$); see e.g. \cite{Dalziel1991Two-layerApproach, Zaremba2003FrictionalFlow, Gu2005} for details. In such a case, the flow is said to be hydraulically controlled. However, enforcing the composite Froude number to be equal to unity at the edges \CB{is inconsistent with the results of the perturbation analysis leading to the HGV-A approximation. On the one side, for internal hydraulics to be valid, inertia in the $x$-component of the momentum equation should not be negligible. On the other hand, setting the value of the composite Froude number at $x=\pm1$ would mean that the position of the interface at these locations depends on the Froude number. Imposing this boundary condition to find the value of $K$ would, in turn, make $K$ depend on the Froude number. However, this is inconsistent with the results from the asymptotic analysis that $K=O_s(A^0)$ while $\Fr=O_s(A^{-1})$, which is a condition needed for the hydrostatic balance \eqref{eq:ns2dz_nd2} to hold.}
\\ 
We propose then a different boundary condition inspired in the results by \cite{Kaptein2020} for the simulation used to exemplify the HGV-A approximation  (\CB{$\Reg=500$}, $A=60$, $\Sc=300$, $\theta=0^\circ$) and shown in figure \ref{fig:rho_sim}. In this figure, we can see that the \CB{interface curves sharply} when reaching the end of the duct, with the light fluid turning upwards at $x=-1$ and the dense fluid turning downwards at $x=1$. \CB{The currents at the edges of the duct must have a dimensionless thickness $\epsilon$ in the $x$-direction with $\epsilon<A^{-1}\ll1$, meaning that it is thin with respect to the length of the duct. We propose then to impose $\eta[x=\pm( 1+\epsilon)]=\mp 1$. For long ducts, we can assume a small error of order $A^{-1}$, and enforce, instead $\eta(x=\pm 1)=\mp 1$, meaning also that $d\eta/dx|_{x=\pm1}=\mp \infty$. Notice that, even though the flow inside the duct is not controlled by the composite Froude number being equal to one at the edges, the boundary conditions at the edges remain crucial in determining the slope of the interface, and hence, the volume flow rate. Taking $\eta(\pm1)=\mp 1$ yields $K\approx 131$ and $d\eta/d x |_{x=0}=-48/K \approx-0.366$ for a horizontal duct. Notice that the value of $d\eta/d x |_{x=0}$ is close to the slope of $-1/3$ obtained empirically and numerically by \citet{Kaptein2020} and shows good agreement with the density field shown in figure \ref{fig:rho_sim}.}

\begin{figure}
    \centering
    \includegraphics[width=0.6\textwidth]{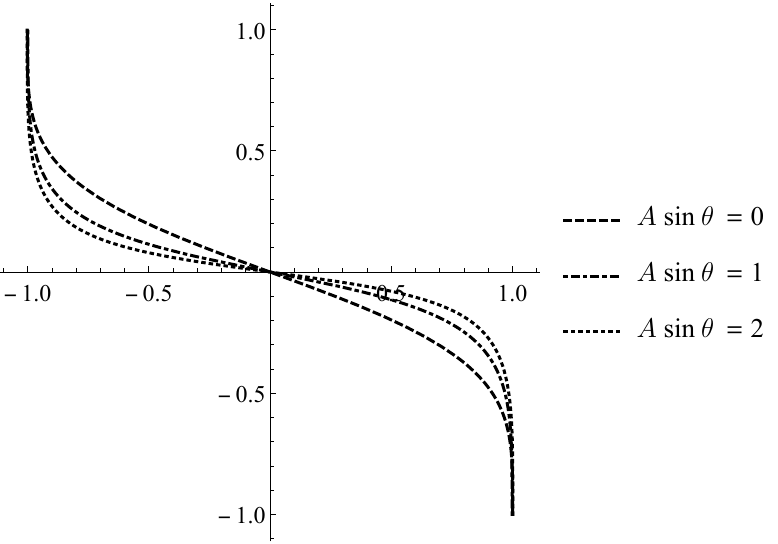}
    \caption{Shape of the interface for three different values of $A\sin\theta$ as obtained from solving \eqref{eq:detadx}. }
    \label{fig:analytical_int}
\end{figure}

\begin{figure}
    \centering
    \includegraphics[width=0.49\textwidth]{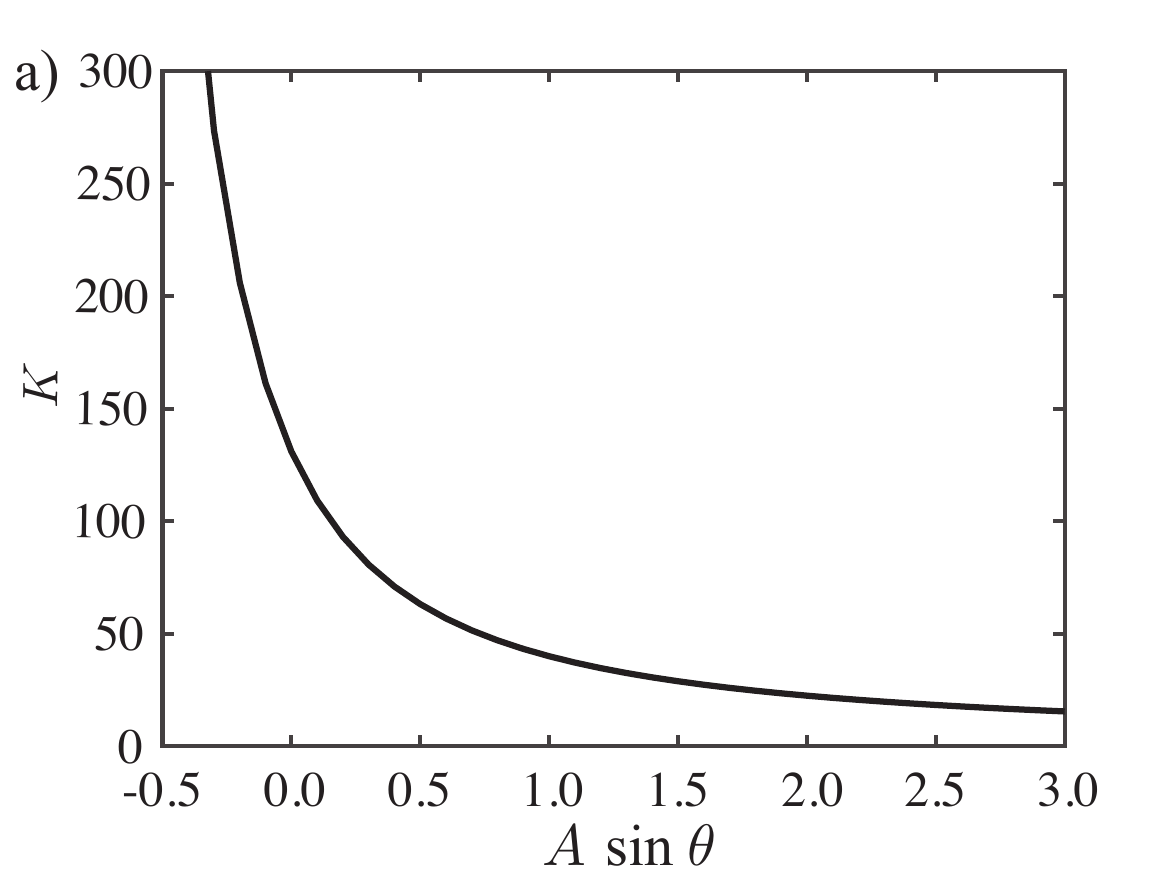}
    \includegraphics[width=0.49\textwidth]{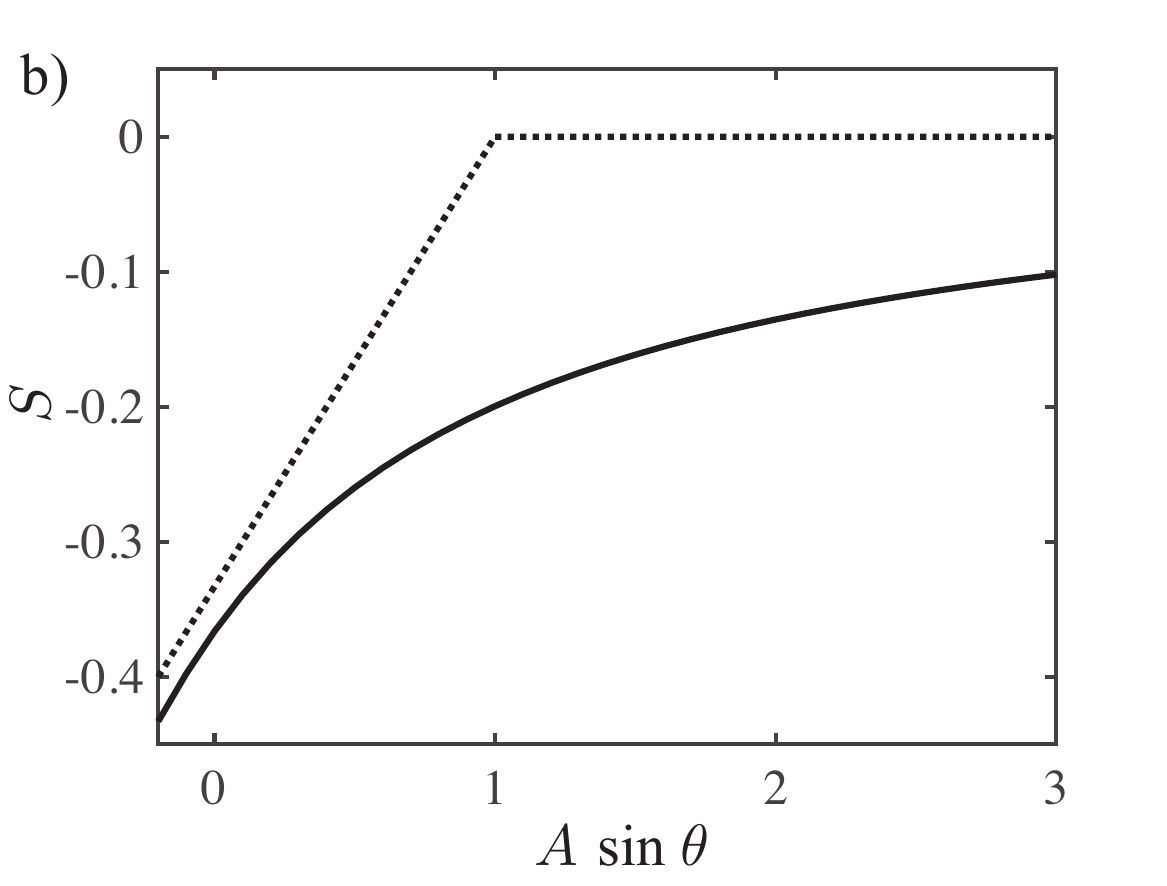}
    \caption{a) Value of \CB{the parameter $K$ defined in \eqref{eq:K}} and b) value $S$, the slope of the interface at $x=0$, as a function of $A \sin \theta$. The value of $K$ and $S$ are such that the solution to the autonomous equation for $\eta(x)$ \eqref{eq:detadx} satisfies $\eta(\pm1)=\mp1$ assuming $\cos \theta \approx 1$. \CB{The dotted line in b) respresent the empirical approximation used by \citet{Kaptein2021ReproducingDomains}.}}
    \label{fig:K_slope}
\end{figure}

\CB{To find the value of $K$ for the inclined ducts, we follow a similar approach as for horizontal ducts, but we do the calculations numerically. The value of $K$ is obtained by solving \eqref{eq:detadx} while enforcing $\eta(\pm 1)=\mp 1$. We further simplify the problem by taking $\cos\theta\approx 1$ since $\theta\ll1$. It was already discussed in Section \ref{sec:asymptotic} that $K$ is a constant for $A^{-1}\to 0$ but that it depends on $A \sin \theta$, and this can be seen again in \eqref{eq:detadx}.} The solution to \eqref{eq:detadx} for three different values of $A \sin \theta$ is shown in figure~\ref{fig:analytical_int}. Several properties of the shape of the interface expected from previous work \citep[see e.g.][]{Gu2005,Lefauve2019,Kaptein2020} and observed in figure \ref{fig:rho_sim} are reproduced by the analytical solution. For example, the solution naturally yields a slope of the interface that is constant to a good approximation over a large portion of the duct around $x=0$, and that it bends up or down as it approaches the edges of the duct. Furthermore, the slope of the interface at $x=0$ given by \CB{
\begin{equation}\label{eq:slope}
       S\equiv \dfrac{d \eta(x)}{d x}|_{x=0}\approx\cos \theta\dfrac{d \eta(x)}{d x}|_{x=0}=-\dfrac{48}{K}+A \sin\theta
\end{equation}}
decreases for increasing values of $\theta$. The values of $K$ and $S$ as a function of $A\sin\theta$ are shown in figure~\ref{fig:K_slope}. \CB{Notice that, due to the way the variables where made dimensionless, the `real' slope (e.g. as observed in the experiments) is given by $S'=S/A$.} 

\CB{Previous work has suggested a change in behavior around $A\sin \theta =1$ where the flow transitions from \emph{lazy} to \emph{forced} \citep{Lefauve2019}. In particular, the slope of the interface is considered relatively flat throughout the duct ($S\approx 0$) for $A \sin \theta >1$. Although here $S$ does not tend to zero for $A \sin \theta >1$, the values of $S'$ are quite small for typical values of $A$ used in experiments. For example, in the case of $A=30$, the height of the interface varies about 3 mm over 1 m, which could be imperceptible by eye. Furthermore, we can see that for $A \sin \theta >1$, the value of $K$ varies little when compared to variation for $A \sin \theta <1$.}

\CB{The fact that we obtain the value of $S$ as a function of $A \sin\theta$ is a critical difference with respect to the parametrization used by \cite{Kaptein2021ReproducingDomains} to determine the regime transition curves. In that work, the variation of $S$ with $A \sin \theta$ was assumed based on previous results by \citet{Lefauve2019} and \citet{Kaptein2020}. In particular, it was assumed that $S= (A \sin \theta-1)/3$ for $A\sin\theta<1$ and that $S=0$ for $A\sin\theta\geq 1$. Although certain features and trends of the regime transition curves might be reproduced using these assumptions, differences are also expected when the value of $S$, as shown in figure~\ref{fig:K_slope}b, is considered.}

\subsection{Implications of the HGV-A approximation for the regime transition}\label{sec:implications}

The derivation of the analytical solution in the HGV-A approximation yielded that $K$ is a constant for a given value of $A\sin\theta$ by solving \eqref{eq:detadx} while enforcing $\eta(\pm 1)=\mp 1$. For the upcoming discussion and an easier comparison with the work by \citet{Lefauve2019}, it is convenient to use the fact that, for a long duct with small tilt angle ($\theta,A^{-1}\ll1$), and $A\sin \theta\approx \theta/\alpha$. In such a case, the solution in the HGV-A approximation yields that $K = \Rey_g/ (\Fr A)$ must be a constant for a given value of $\theta/\alpha$. Both $\Rey_g$ and $A$ are control parameters of the problem, while $\Fr$ is a response parameter equivalent to the non-dimensional volume flow rate. Hence, if the value of $\Rey_g A^{-1}$ is increased while keeping $\theta/\alpha$ fixed, the value of $\Fr$ should also increase keeping the value of $K$ constant. \\
\CB{It is, here, convenient to define the Froude number in the HGV-A approximation: 
\begin{equation}
    \Fr^* \equiv \frac{\Rey_g }{A K}\approx \frac{\Rey_g \alpha}{K} .
\end{equation}
Note that $\Fr^*$ can be seen as either a response parameter or a control parameter. It is a response parameter because it represents the non-dimensional volumetric flow rate which is a results of the choice of the control parameters: $\Rey_g$, $\theta$, and $A$. In addition, it is a control parameter because $\Reg$ and $A$ are directly imposed for a given experiment, and $K$ is a geometrical parameter set by imposing the value of $\theta/\alpha$. The value of $K$ is determined from the analytical solution in the HGV-A approximation, and it is, to a very good approximation for $\theta\ll 1$, the value shown in Figure \ref{fig:K_slope}a. In this way, the value of $\Fr^*$ is a known quantity for a given experiment. Within the HGV-A approximation, $\Fr = \Fr^*$, but this is not the case if the approximation does not hold. Furthermore, since $K\approx 131$ is a constant for $\theta = 0$, saying that $\Fr^*$ is the control parameter for horizontal ducts is equivalent to saying that $\Rey_g\,A^{-1}$ is the control parameter as shown by \citet{Hogg2001}.} 

\CB{To study the limit of validity of the HGV-A approximation, we consider a hypothetical SID setup with given values for $A$ and $\theta$ satisfying $A\gg 1$ and $\theta\ll 1$. As just mentioned, the value of $K$ is known then. In this hypothetical setup, we first take a sufficiently small value of $Re_g$ so that $\Fr^*$ is sufficiently small to  neglect inertia from the $x$-component of the momentum equation \eqref{eq:ns2dx_nd}. Furthermore, $Sc\gg 1$ so that the flow gets organized (too good approximation) in a two-layer configuration and the HGV-A approximation holds. In such a case, $Fr=\Fr^*$. We now do a series of experiments increasing $\Rey_g$, which is equivalent to increasing $\Fr^*$. We know from internal hydraulic theory that there is a maximum flow rate possible through the duct, i.e. a maximum possible value for $Fr$. In the frictionless case, this maximum value is $Fr=1$ which is known as the \emph{hydraulic limit} \citep{Hogg2001,Lefauve2019}, while friction reduces the maximum possible value of $Fr$ \citep{Gu2005}. Hence, there must be a critical value $\Fr^*=\Fr^*_c$ so that $\Fr^*>\Fr$ when $\Fr^*>\Fr^*_c$. If this is indeed the case, the HGV-A approximation is invalid $\Fr^*>\Fr^*_c$. Hence, the limit in the validity of the HGV-A approximation should be given by $\Fr^*=\Fr^*_c$, where it is expected that a transition (i.e. a qualitative change in the flow) occurs.}


\CB{Now, the next step is to determine the range of validity of the HGV-A approximation by determining $\Fr_c$ since the asymptotic analysis does not give information about the physical order of magnitude at which the different terms become relevant \citep{VanDyke1975PerturbationMechanics}. Hence, to determine the range of validity of the HGV-A approximation, it is necessary to use numerical simulations or laboratory experiments. In the following section, we study this range of validity and the consequence for regime transitions in four different experimental setups.}\\

\section{Experimental verification}\label{sec:data}

It was already discussed that the several of the properties of the analytical solution in the HGV-A approximation agree with the numerical results of \cite{Kaptein2020}. Now, we use  
results from laboratory experiments to verify the theoretical results derived in the previous section. We have two main aims: $i$) show that the solution in the HGV-A approximation described in Section \ref{sec:analytic} does exist and is observed experimentally in a region of the parameter space, $ii$) show that curves of \CB{$\Fr^*=\text{const.}$} describe the transition between different regimes. 

\subsection{Description of the data}
We use the experimental data sets by \citet{Meyer2014}, \citet{Lefauve2019} and \citet{Lefauve2020}. These data sets have been discussed and made available online (https://doi.org/10.17863/CAM.48821, https://doi.org/10.17863/CAM.41410) by \citet{Lefauve2019b} and \citet{Lefauve2020,Lefauve2020a}. Four ducts with different dimensions were used. We will refer to them as mSID, tSID, LSID and HSID in accordance with \citet{Lefauve2020}. Table \ref{tab:table_exp} summarizes the characteristics of all the setups. For each setup, the inclination angle of the duct $\theta$ and the gravitational Reynolds number $\Rey_g$ were varied independently. The fluid used was a salt (NaCl) solution ($Sc\approx700$), and the value of $\Rey_g$ was varied by changing the salt concentration in each of the tanks. In total, we use 738 data points to study the regime transitions. 

\begin{table}
    \centering
    \begin{tabular}{lccccccc}
        name & $H$ (mm) & Cross-section & $A$ & $B$ & $Sc$ & $\theta$ (deg.) & $\Reg$\\
        mSID & 45 & \begin{tikzpicture} \draw (0,0) -- (0.4,0) -- (0.4,0.4) -- (0,0.4) -- (0,0); \end{tikzpicture} & 30 & 1 & 700 & [-1,6] & [300, 6 000]\\
        LSID & 100 & \begin{tikzpicture} \draw (0,0) -- (0.8,0) -- (0.8,0.8) -- (0,0.8) -- (0,0); \end{tikzpicture} & 30 & 1 & 700 & [-1,4] & [2 000, 20 000]\\
        HSID & 100 & \begin{tikzpicture} \draw (0,0) -- (0.8,0) -- (0.8,0.8) -- (0,0.8) -- (0,0); \end{tikzpicture} & 15 & 1 & 700 & [0,4] & [1 000, 20 000]\\
                tSID & 90 & \begin{tikzpicture} \draw (0,0) -- (0.2,0) -- (0.2,0.8) -- (0,0.8) -- (0,0); \end{tikzpicture} & 15 & 1/4 & 700 & [-1,3] & [3 000, 15 000]\\
    \end{tabular}
    \caption{Characteristics of the experiments used in this paper. Four duct geometries [abbreviated mSID (m for mini), LSID (L for large), HSID (H for half), tSID (t for tall)] are used \citep{Lefauve2020}. We list the values of the dimensionless numbers describing each duct geometry ($A$ and $B$), the value of $Sc$ for salt in water, and the ranges of $\theta$ and $\Reg$ explored.}
    \label{tab:table_exp}
\end{table}

\citet{Meyer2014} distinguished four different regimes: laminar ($\mathsf{L}$), Holmboe waves ($\mathsf{H}$), intermittently turbulent ($\mathsf{I}$), and turbulent ($\mathsf{T}$). \citet{Lefauve2020} introduced a `waves' ($\mathsf{W}$) regime where waves other than Holmboe waves were observed. The different regimes were mostly identified by shadowgraph observations over a subsection of the duct, following the qualitative description of each regime by \citet{Meyer2014}. A schematic of the shadowgraph set-up was presented by \CB{\citet{Lefauve2018b}}. The observed regime and the mass flow rate as a function of the governing parameters are provided by \cite{Lefauve2020a}. A complication for the comparison with the experimental data is that, as pointed out by \citet{Lefauve2020}, there is a surprising difference between the results from the LSID and the mSID setups: the regions in the $(\theta,\Reg)$-plane where the different regimes occur do not coincide even though the values of all the dimensionless parameters are identical.

\CB{To show that the proposed solution in the HGV-A approximation is observed experimentally and to observe the limit of validity, we analyse in the following section the along-duct velocity and the density fields for three experiments in the mSID duct all with $\theta =2^\circ$ \citep[provided by][]{Lefauve2019b}. The first experiment falls within the $\mathsf{L}$ regime with $\Reg=398$, the second within the $\mathsf{H}$ regime with $\Reg=1059$, and the third within the $\mathsf{I}$ regime with $\Reg=1466$. A detailed description of the experiments and the methodology is given by \citet{Lefauve2019}. With this choice of experiments, we have the same approach as mentioned for a hypothetical SID setup in section \ref{sec:implications}, but with a real setup and real experimental results.}

\subsection{Experimental confirmation of the HGV-A approximation}\label{sec:confimation}
\begin{figure}
    \centering
    \includegraphics[width=1.0\textwidth]{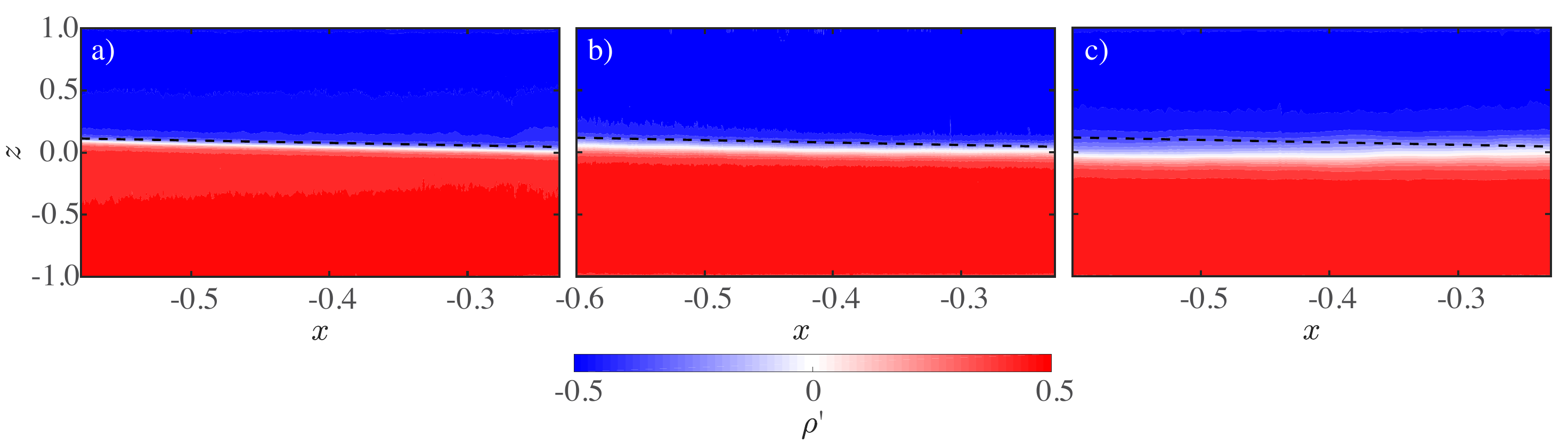}
    \caption{\CB{Time-averaged density field in the $(x,z)$-plane at $y=0$ for three experiments in the mSID setup. a) Experiment within the $\mathsf{L}$ regime ($\theta =2^\circ$, $\Reg=398$). b) Experiment within the $\mathsf{H}$ regime ($\theta =2^\circ$, $\Reg=1059$). c) Experiment within the $\mathsf{I}$ regime ($\theta =2^\circ$, $\Reg=1455$). The black dashed line represents the line $z=Sx$  with $S=-0.20$ as given by \eqref{eq:slope}.}
    }
    \label{fig:density}
\end{figure}

\CB{For the three experiments for which we analyze the velocity and density field, we first determine the value of $K$ numerically as explained in Section \ref{sec:K}. This value is given to good approximation in figure \ref{fig:K_slope}a. For  $\theta = 2^\circ $ and $A=30$, $A \sin \theta \approx \theta/\alpha \approx 1.0$ so $K\approx 39$. We then compute the value of the Froude number in the HGV-A approximation $Fr^*\equiv\Reg (K\,A)^{-1}$, giving $\Fr^*\approx 0.34$ for the $\mathsf{L}$ experiment, $\Fr^*\approx0.91$ for the $\mathsf{H}$ experiment, and $\Fr^*\approx 1.25$ for the  $\mathsf{I}$ experiment. Here, we notice already that the value of $\Fr^*$ for the $\mathsf{I}$ experiments is larger than $\Fr=1$ corresponding to the hydraulic limit. In the following, we investigate if the key flow characteristics and working assumptions leading to the solution in the HGV-A approximation derived in section \ref{sec:analytic} hold in these experiments. We consider four aspects: (i) the density field, focusing on the slope of the interface; (ii) the $x$-component of the momentum equation \eqref{eq:upsilon_nl}, focusing on the fact that the HGV balance holds; (iii) the the vertical profiles of the along-channel velocity component; and finally, (iv) the prediction of the mass flux through the duct.}

\CB{Figure \ref{fig:density} shows the experimentally obtained time-averaged density fields and the location of the interface as predicted by the HGV-A approximation for the three individual experiments considered. The slope of the interface close to $x=0$ is given to good approximation by \eqref{eq:slope}, which yields $S\approx -0.20$ for these experiments. First, we see that the density is indeed organized into two layers with a sharp interface in between, with the $\mathsf{I}$ experiment presenting a thicker interface. For the $\mathsf{L}$ experiment, the semi-analytical solution for the slope of the interface agrees well with the experiment. This means that \eqref{eq:detadx} with the boundary condition $\eta(\pm 1)=\mp 1$ predicts well the shape of the interface for this experiment and that the value of $K$ obtained from the analytical solution in the HGV-A approximation applies. The agreement is less for the $\mathsf{H}$ experiment, with the slope of the interface being slightly less steep than predicted. Finally, for the $\mathsf{I}$ experiment, the interface is fully parallel to the top and bottom of the duct, suggesting that the solution in the HGV-A approximation does not hold anymore for this experiment.}

\begin{figure}
    \centering
    \includegraphics[width=1.0\textwidth]{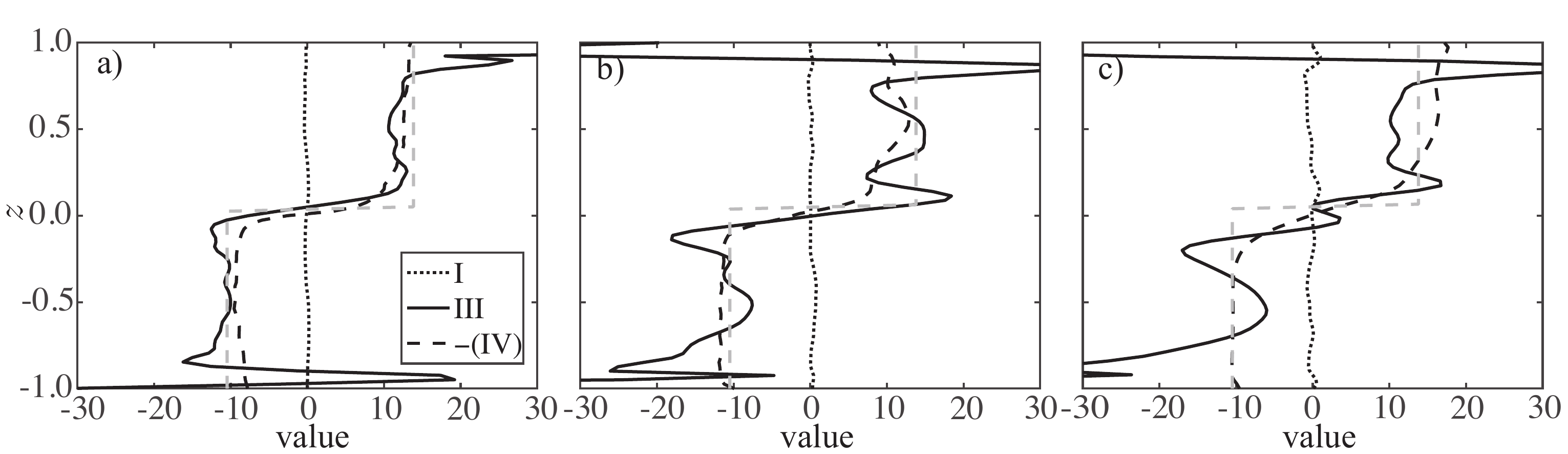}
    \caption{\CB{The vertical profiles of the time- and $y$-averaged terms of the horizontal momentum balance \eqref{eq:upsilon_nl} as a function of $z$ and $x\approx -0.2$ for the same three experiments as in figure \ref{fig:density}. The black dotted, solid and dashed lines denote the terms I, III, and (minus) IV, respectively, as obtained from the laboratory experiments. Term II is always approximately equal to zero and not explicitly shown. The dashed grey line represents both terms III and (minus) IV as obtained from the analytic solution in the HGV-A approximation which assumes an infinitely sharp interface, and it is given by the right hand side of \eqref{eq:HGV-A_u}.} }
    \label{fig:equation}
\end{figure}

\CB{As a next step, we investigate if the HGV balance \eqref{eq:HGV-A_u} is indeed the leading order balance in the $x$-component of the momentum equation \eqref{eq:ns2dx_nd} close to the center of the duct ($x\approx 0$), which we rewrite here as 
\begin{equation}\label{eq:upsilon_nl}
    \underbrace{A^{-1} \Rey_g \Fr  \left(u\frac{\partial u}{\partial x}+w\frac{\partial u}{\partial z}\right)}_{I}  = 
\underbrace{A^{-2} \frac{\partial^{2} u}{\partial x^2}}_{II}+\underbrace{\frac{\partial^{2} u}{\partial z^2}}_{III}+\underbrace{\dfrac{\Rey_g }{2A\Fr}\left( A \sin \theta\rho'-\cos\theta \int_{0}^{z}\dfrac{\partial\rho'}{\partial x}dz'\right)}_{IV},
\end{equation}
with $z'$ a dummy variable. Here, we have assumed that the hydrostatic balance \eqref{eq:ns2dz_nd2} holds and followed the derivation by \citet{Kaptein2020} to write the pressure gradient term in this form. To see that the HGV balance \eqref{eq:HGV-A_u}, vertical viscous diffusion (term III) must be equal to minus the forcing (term IV). In addition, we want to see that term I (inertia) is negligible with respect to term III and term IV. Figure \ref{fig:equation} shows with black lines the vertical profiles of terms I, III and (minus) IV of \eqref{eq:upsilon_nl} obtained from the time- and y-averaged velocity and density fields at $x\approx -0.2$. We do not show term II because it is practically equal to zero for all experiments. In general, we see that term I is also much smaller than the other two terms meaning that it can indeed be neglected. As can be seen, term III and (minus) term IV overlap well for all experiments, with the experiment in the $\mathsf{L}$ regime showing better agreement, meaning that the HGV balance holds particularly well for the $\mathsf{L}$ experiment. In the case of the $\mathsf{H}$ and $\mathsf{I}$ experiments, terms III and IV match in general, suggesting that the HGV balance is still the leading balance of the time-averaged flow, but some deviations are observed.}

\CB{To understand these deviations, we make a comparison with the theory. Figure \ref{fig:equation} also shows with dashed grey lines the value for terms III and IV according to the solution in the HGV-A approximation derived in section \ref{sec:analytic} and given by the right hand side of \eqref{eq:HGV-A_u}, which assumes an infinitely thin interface. The interface in the experiments is not infinitely thin due to diffusion in the vertical direction which is fully neglected in the HGV-A approximation. This effect is most visible in the $\mathsf{I}$ experiment due to turbulent diffusion across the interface. For all three experiments, the forcing corresponds well with forcing expected in the HGV-A approximation given by the sum of gravity and the baroclinic pressure gradient. It is the diffusion in the vertical direction that shows deviations for the $\mathsf{H}$ and $\mathsf{I}$ experiments. We notice, in particular, an increase of the magnitude of vertical momentum diffusion in the bottom boundary layer, and just above and below the shear layer. Notice that the reasons why the momentum balance is not closed for the $\mathsf{H}$ and $\mathsf{I}$ experiments is that \eqref{eq:upsilon_nl} would need additional terms for the time-averaged experimental results. Most importantly, it misses the momentum diffusion in the across-duct direction and a term containing the time-averaged effect of temporal perturbation (similar to Reynolds stresses).}

\begin{figure}
    \centering
    \includegraphics[width=1.0\textwidth]{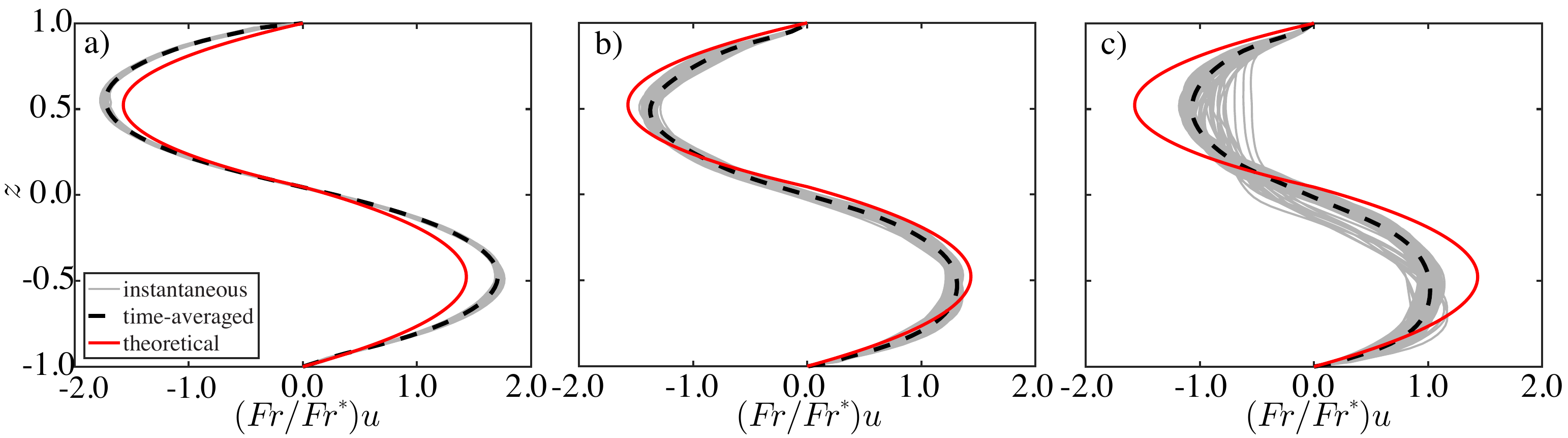}
    \caption{\CB{Vertical velocity profiles of the $x$-component of the velocity as a function of $z$ for $x\approx -0.2$ and averaged in the $y$-direction for the same three experiments as in figures \ref{fig:density} and \ref{fig:equation}. The grey lines represent the instantaneous velocity, the black dashed line the time-averaged velocity, and the red line represents the velocity in the HGV-A approximation given by \eqref{eq:u_HGV-A_F}.} }
    \label{fig:velocity}
\end{figure}

\CB{We now compare the velocity profiles. Figure \ref{fig:velocity} shows the vertical profiles of the $x$-component of the velocity at $x\approx -0.2$ and averaged in the $y$-direction for the same three experiments. Choosing different values of $x$ does not affect the results beside the fact that the height of the point with $u=0$ is slightly shifted. For the experiment in the $\mathsf{L}$ regime, the flow is steady so the instantaneous and time-averaged velocity profiles coincide. Even for the $\mathsf{H}$ and $\mathsf{I}$ experiments, the variation in the instantaneous profiles is small compared to the magnitude of the time-averaged velocity. For the three experiments, we see a different level of disagreement in the amplitude of the velocity. Computing the value of $\Fr$ for the experiments using the time- and $y$-averaged $x$-velocity component, yields $\Fr=0.40$ (compared with $\Fr^*\approx 0.34$) for the $\mathsf{L}$ experiment, $\Fr=0.81$ (compared with $\Fr^*\approx 0.91$) for the experiment in the $\mathsf{H}$ regime, and $\Fr=0.90$ (compared with $\Fr^*\approx 1.25$) for the experiment in the $\mathsf{I}$ regime. As mentioned earlier, maximum exchange theory says that the maximum value of $\Fr$ is $\Fr=1$, so we could have already expected that the Froude number for the $\mathsf{I}$ experiment was not going to reach $\Fr^*=1.25$, but we have here confirmation of this. This suggests again that the HGV-A approximation does not hold for this experiment. On the other hand, the value of the Froude number for the $\mathsf{L}$ and $\mathsf{H}$ experiments are well predicted (within 13\%) by the HGV-A analytical solution. We believe that this is a very good result, particularly, if we consider, on one side, experimental errors, and on the other, some of the assumptions made during the analytical derivation. For example, we assumed an infinitely wide duct, while the experiments have a cross-sectional aspect ratio $B=1$.}

\CB{In the previous paragraphs, we studied the characteristics of only three experiments with equal value of $\theta/\alpha$, and hence, equal value of $K$. In general, we saw that the $\mathsf{L}$ experiment is very well reproduced by the HGV-A approximation, the $\mathsf{H}$ is slightly less so, and the $\mathsf{I}$ shows clear differences.  In section \ref{sec:implications}, we discussed that we expect a critical value $\Fr^*_c$ such that $\Fr<\Fr^*$ when $\Fr^*>\Fr^*_c$. The results suggest that $\Fr^*_c$ is found somewhere between the $\mathsf{H}$ and the the $\mathsf{I}$ experiment. However, it is difficult to determine clear trends from only these three experiments. Although values of $Fr$ are not available for many experiments, the mass flux 
\begin{equation}
Q_m\equiv\int_{-1}^{1}\int_{-1}^{1}|\rho' u|dydz
\end{equation}
is available for several experiments. The value of $Q_m$ is commonly used to compare (and thus characterize) the flow, with $Q_m=1/2$ corresponding to the hydraulic limit \citep{Hogg2001,Lefauve2019}, while $Q_m=\Fr/2=\Fr^*/2$ in the HGV-A approximation. The factor $1/2$ arises from the fact that $\rho'=\pm 1/2$ in the HGV-A approximation following the way we made the density dimensionless.}

\begin{figure}
    \centering
    \includegraphics[width=0.47\textwidth]{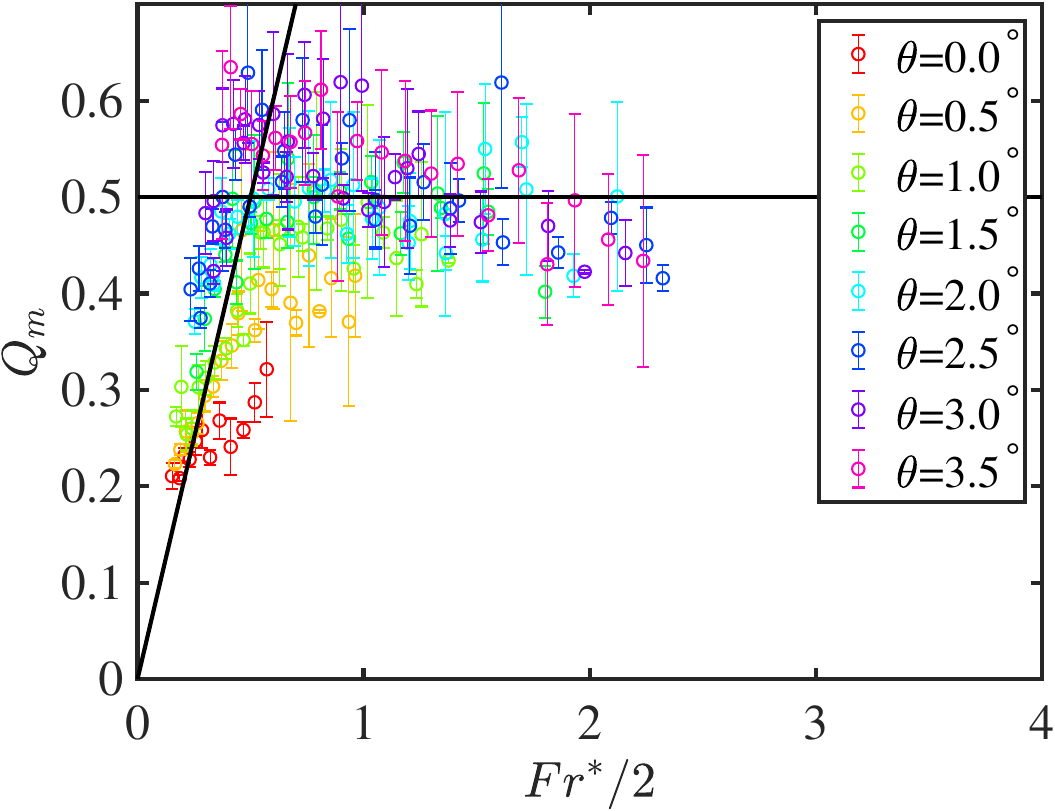}
    \caption{Mass flux per unit width $Q_m$ as a function of the parameter $\Fr^*/2$ for the experiments in the mSID setup. The color denotes the value of the angle $\theta$. The solid black lines represent $Q_m=\Fr^*/2$ (the expected value in the HGV-A approximation) and $Q_m=1/2$ (the expected value in the hydraulic limit). \CB{The experimental values tend to follow the trend of the prediction based on the HGV-A approximation $Q_m\approx Fr^*/2$ until $\Fr^*/2\approx 0.5$. For larger values of $\Fr^*/2$, a maximum, constant value of $Q_m$ is reached for each value of $\theta$ as expected from (frictional) two-layer hydraulics. The overall maximum value is $Q_m\approx 0.5$ as predicted for the hydraulic limit.}}
    \label{fig:Q_m}
\end{figure}

The value of $Q_m$ as a function $\Fr^*/2$ for the experiments in the mSID setup are shown in figure \ref{fig:Q_m}. The values of $Q_m$ can then be compared to the expected values in the HGV-A approximation ($Q_m=\Fr^*/2$) and to the hydraulic limit ($Q_m=1/2$). For $\Fr^*/2\lesssim 1/2$, $Q_m\approx\Fr^*/2 $, showing that the HGV-A approximation gives a good estimate for the value of $Q_m$. Although the values for some specific experiments are overestimated, and some other are underestimated, clear trends are visible. For example, for the experiments with $\theta\geq 1.0^\circ$, $Q_m\approx\Fr^*/2$ up to $\Fr^*/2=1/2$. Then, a sharp transition to the hydraulic limit occurs, and $Q_m\approx 1/2$ for $\Fr^*/2\geq 1/2$. For experiments with $\theta = 0.5^\circ$, $Q_m$ seems to saturate reaching a maximum value $Q_m\approx 0.4<1/2$. The results for $\theta = 0^\circ$ suggest that $Q_m$ will similarly not reach the value $1/2$ for a horizontal duct. The impossibility of reaching $Q_m=1/2$ is in agreement with frictional hydraulic theory which states that frictional effects reduce the maximum possible value of $Q_m$ \citep{Gu2005}. It is only for $\theta\geq1.0^\circ$ that the additional forcing by gravity in the along-duct direction reduces the importance of frictional effects and makes the direct transition from the HGV-A approximation to the hydraulic limit ($Q_m=1/2$) possible.

\CB{Figure 8 is reminiscent of similar plots by \cite{Hogg2001} where they plotted $Q_m$ as a function of $(\Rey_g A^{-1})^2$ for simulations with $Sc=1$ in a horizontal duct. In that case, they observed the change from the VAD solution to the hydraulic limit. In our case with $Sc\approx 700$, we observe a similar change but from the HGV-A approximation to the hydraulic limit. In short, the flow changes from the HGV-A approximation where the value of $K$ is constant for a given value of $\theta/\alpha$, to a hydraulically controlled flow where the value of $\Fr$ is constant for a given value of $\theta/\alpha$.}

In general, the flow in the $\mathsf{L}$ regime is very well described by the HGV-A approximation, and even for the experiment in the $\mathsf{H}$ regime. \CB{For the $\mathsf{L}$ experiment, $\Fr=0.40 $, $\Rey_g=398$ and $A=30$, meaning that $\Fr\Reg A^{-1} \approx 5.3$ or  $Re_g\approx 13.3 A$. We have seen in figure \ref{fig:equation} that inertia can be neglected for this experiment. Furthermore, if we consider the series of experiments with $\theta =1^\circ$ ($K\approx 62$) shown in figure \ref{fig:Q_m}, we see that $Q_m\approx 0.5$ for $\Fr^*/2>0.5$. This suggests that the limit of validity of the HGV-A (in which inertia is neglected) is around $\Fr^*\approx 1$. For this set of experiments, $\Fr^*=1$ corresponds to $\Rey_g\approx 62 A$. These two examples show that the restrictive condition $\Rey_g\ll A $ derived using a scaling analysis by \citet{Lefauve2020} to neglect advection terms can be drastically relaxed.}%

\subsection{Regime transitions}\label{sec:filetypes}

\begin{figure}
    \centering
    \includegraphics[width=0.49\textwidth]{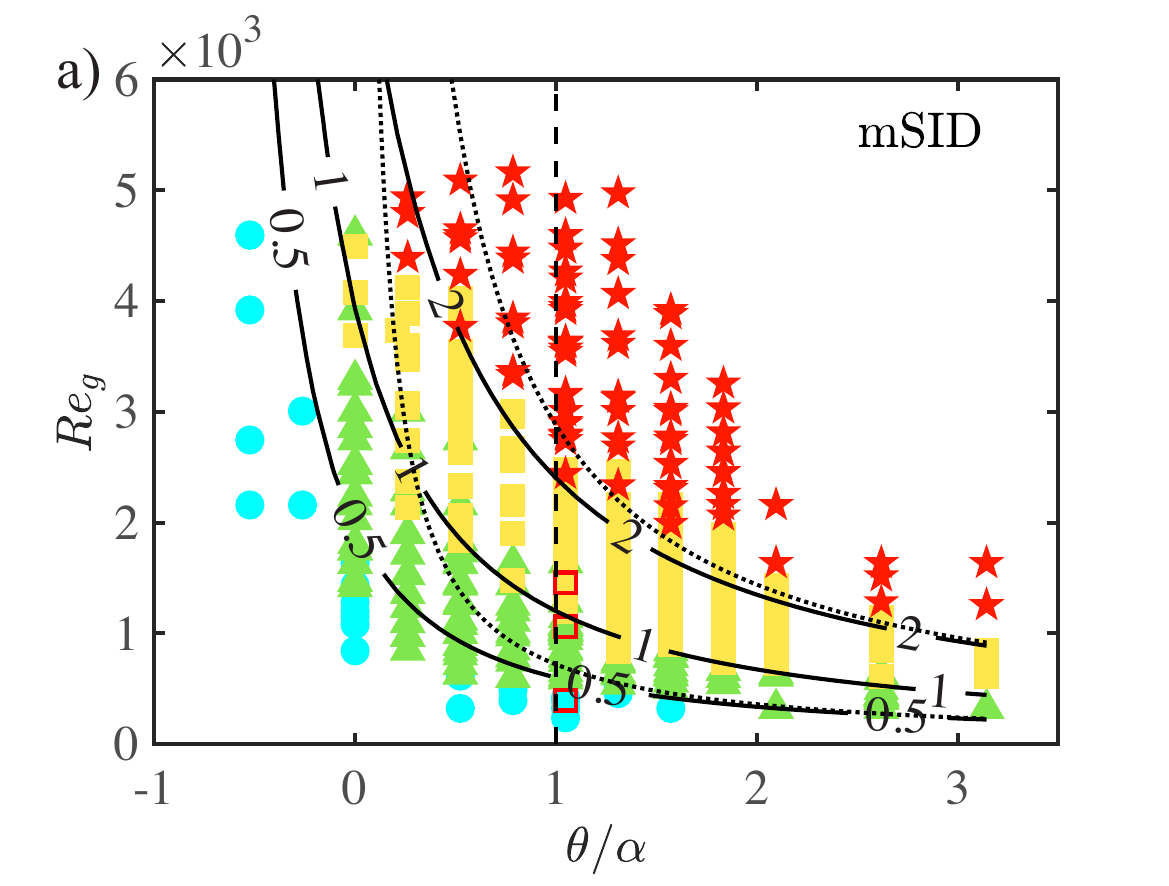}
    \includegraphics[width=0.49\textwidth]{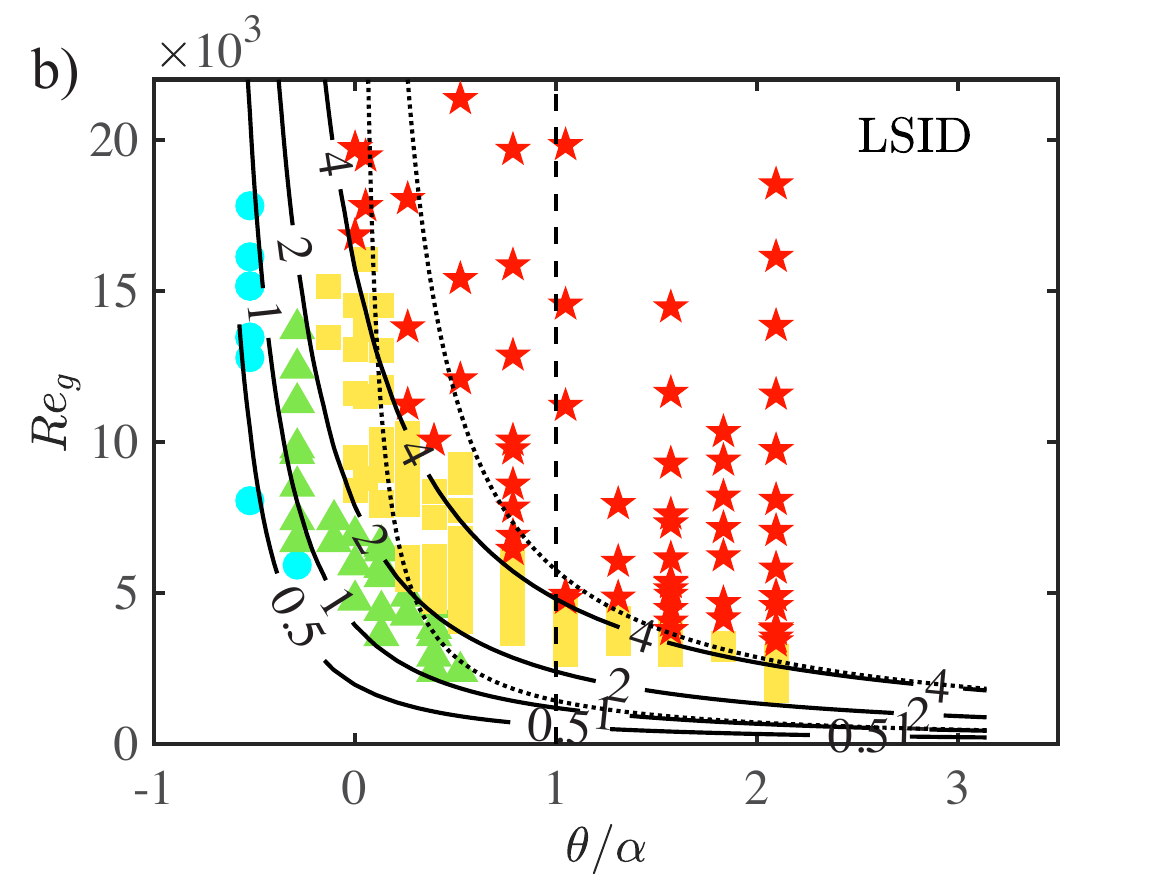}\\
    \includegraphics[width=0.49\textwidth]{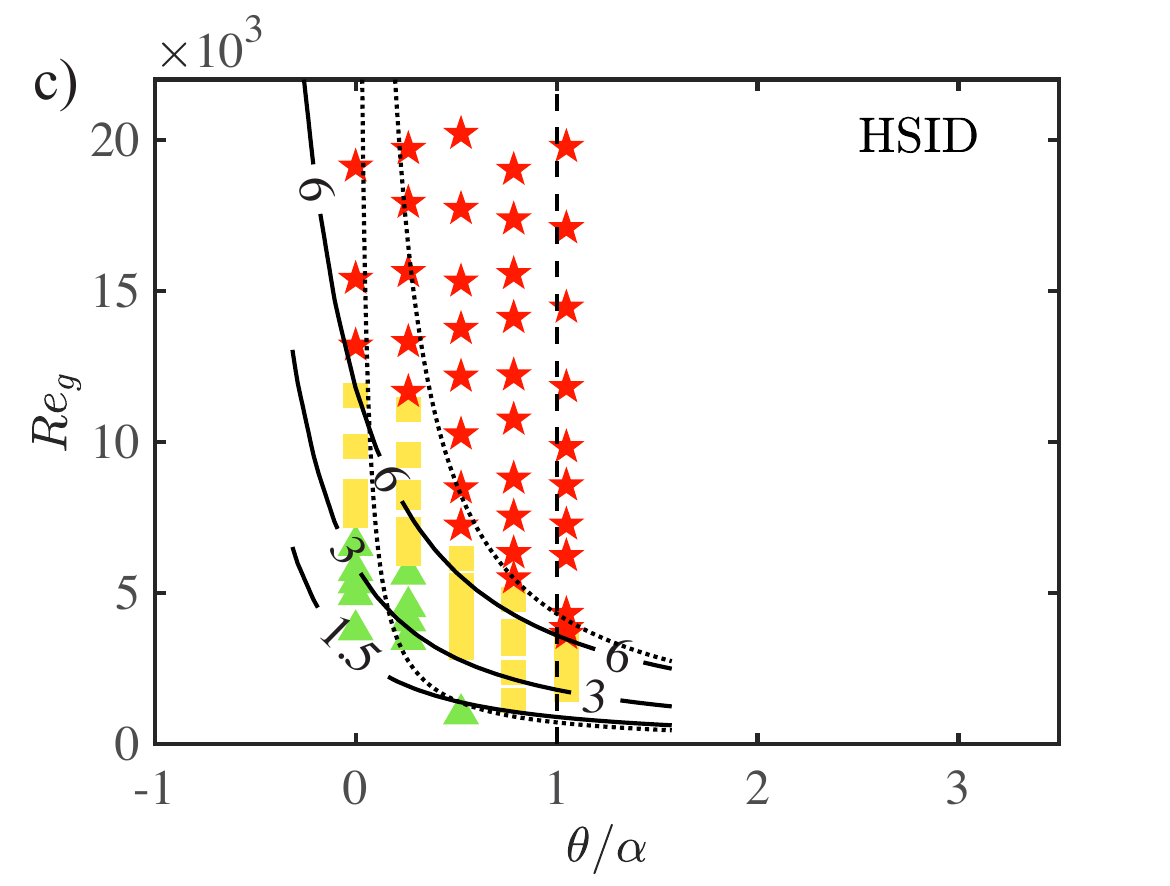}
        \includegraphics[width=0.49\textwidth]{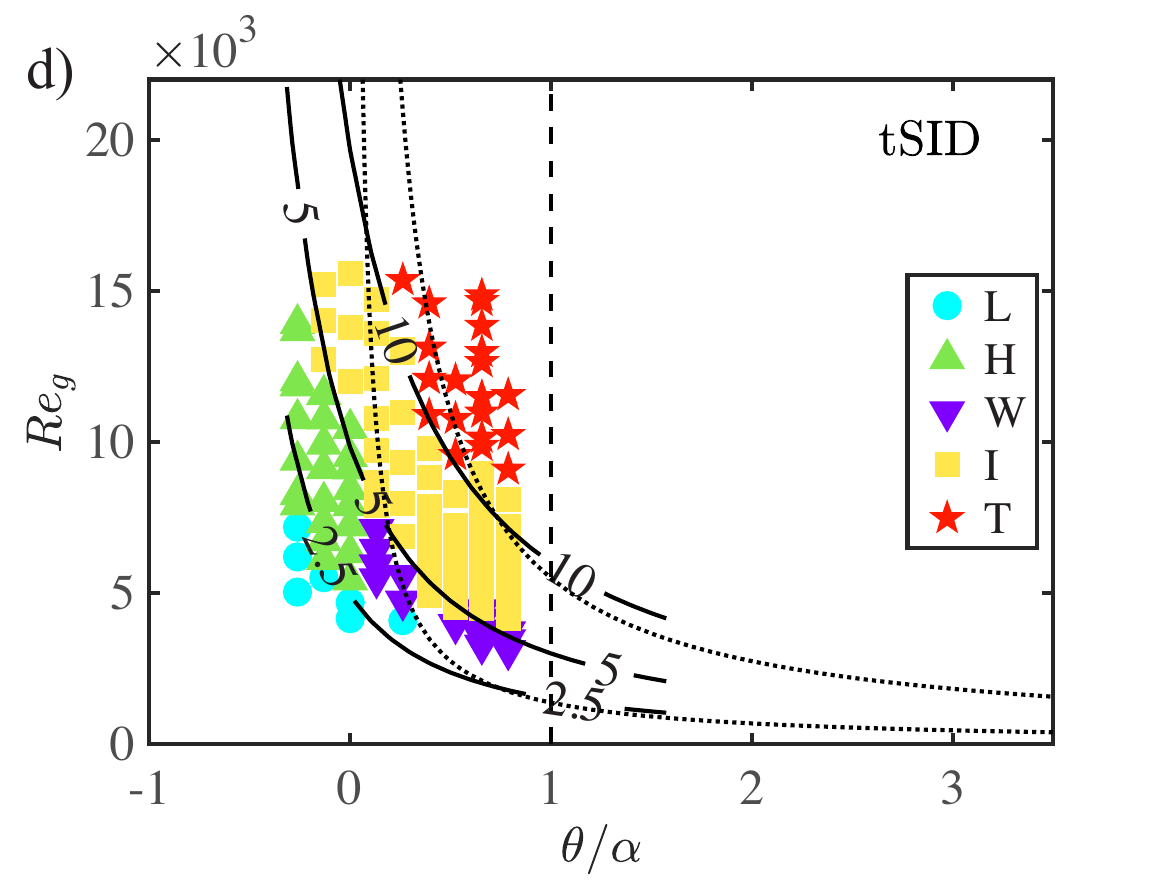}\\
    \caption{Location of the different regimes in the ($\Rey_g$,$\theta$) plane for the four different setups: mSID, LSID, HSID, and tSID. The different symbols represent the different regimes laminar ($\mathsf{L}$), Holmboe waves ($\mathsf{H}$), other waves ($\mathsf{W}$), intermittently turbulent ($\mathsf{I}$) and turbulent ($\mathsf{T}$). The solid lines represent curves of constant $\Fr^*$ with the value indicated along the line. The dashed line represents $\theta/\alpha=1$. The dotted lines are examples of the transition between regimes for \emph{forced} flows given by $\theta\Reg=\text{const.}$ as proposed by \cite{Lefauve2019}.}
    \label{fig:regimes}
\end{figure}

\CB{In this section, we finally discuss how the values of $\Fr^*$ are related to regime transitions.} Figure \ref{fig:regimes} shows the location of the different regimes in the parameter space $(\theta,\Reg)$ for all experimental setups with superimposed curves of constant $\Fr^*$. It can be clearly seen that these curves describe the transitions between the different regimes through all values of $\theta/\alpha$. For $\theta/\alpha \gtrsim 1$,  curves given by $\Fr^*=\text{const.}$ are equivalent to curves given by $\theta \Reg=\text{const.}$ for a given duct. This is in agreement with the transitions derived for \emph{forced} flows by \citet{Lefauve2019} using an energetics budget approach and that \citet{Lefauve2020} derived using a two-layer frictional hydraulics model. However, the agreement of the curves of constant $\Fr^*$-value based on the HGV-A approximation extends to \emph{lazy} flows ($\theta/\alpha<1$) including $\theta/\alpha \leq 0$ where the curves given by $\theta \Reg=\text{const.}$ do not make physical sense.

\CB{In the mSID setup, the transitions between successive regimes are given by $\Fr^*\approx 0.5,$ 1, and 2. The detailed analysis of the three experiments in the previous section (shown as red open squares in figure \ref{fig:regimes}a) and of the mass flux as a function of $\Fr^*$ coincides with this picture. We observe the emergence of turbulence at $\Fr^*\approx 1$, where we find the limit of validity of the HGV-A approximation. This transition further represents a transition to hydraulically controlled flows. It is interesting now to note that experiments in the $\mathsf{I}$ regime (i.e. for $ 1 \lesssim \Fr^* \lesssim 2$) and $\theta \geq 2$ have values of $Q_m$ larger than in the hydraulic limit where $Q_m\approx 1/2$. For these experiments, $Q_m$ seems to overshoot following the trend of the HGV-A approximation ($Q_m=\Fr^*/2$). This suggests that the flow in the $\mathsf{I}$ regime has mixed characteristics of the HGV-A approximation and the fully-turbulent, hydraulically-controlled flow. Finally, the characteristics of the HGV-A approximation are negligible when $\Fr^*\gtrsim 2$ and the flow is in the $\mathsf{T}$ regime.} 

Surprisingly, although curves of constant $\Fr^*$ match the transitions for all ducts, the particular values of $\Fr^*$ associated with the transitions differ. For the LSID setup, the values of $\Fr^*$ marking these transitions are about a factor 2 larger, for the HSID about a factor 3 larger, and for the tSID about a factor 5 larger. In spite of the differences in these values, there is a remarkable similarity: there is always a factor 2 between the values of $\Fr^*$ describing successive regime transitions.


\CB{The difference in the transition values for the tSID setup with respect to all the other setups could have been expected since there is one non-dimensional parameter $B$ which is different. In fact, the tSID setup is much thinner than the other setups, and hence, three-dimensional effects are more important. In particular, there is additional friction from the side walls, which results in a smaller value of $\Fr$ for a given forcing magnitude. This, in turn, means that the successive transitions are delayed. Further investigation of the three-dimensional effects would require to extend the two-dimensional solution presented in section \ref{sec:analytic} to three dimensions, which we believe is possible, but outside the scope of the current paper.}  

\CB{On the other hand, the difference between the mSID and LSID (both with $A=30$), and the HSID (with $A=15$) setups needs further explanation since the HGV-A approximation accounts for having different $A$ values. It could be that $A=15$ is not long enough for the limit of long ducts to hold. However, we do not have enough information to test this hypothesis. Still, the most surprising difference is between the LSID and mSID setups, as already mentioned by \cite{Lefauve2020}, since both setups have the same values for $A$, $B$, and $\Sc$. If the values of these non-dimensional parameters are the same, the fact that two experiments with the same values of $\Rey_g$ and $\theta$ are in different regimes would go against the principle of dynamic similarity. \citet{Lefauve2020} already mentioned that there might be a missing length scale (and its accompanying non-dimensional parameter), but they could not think of a relevant one. If one is to consider just the duct itself, there are no other non-dimensional parameters that are needed to describe the problem. For example, the non-dimensional parameters used so far are the only parameters needed to setup a numerical simulation such as those by \citet{Kaptein2020}. This would suggest that there are certain characteristics of the setup outside the duct that can delay or stimulate flow transitions. Unfortunately, there is much less data available for the LSID and HSID setups when compared to the mSID setup. For example, there are no 3D velocity and density fields, and there are fewer experiments for which the value of $Q_m$ is available. However, we saw that for the mSID setup the different data give a self-consistent picture. Ultimately, new independent experimental or numerical efforts will be needed to clarify the inconsistency between setups.}

\CB{We believe that the key to understand further the regime transitions, and possibly, the differences between the mSID and LSID setups lies at the edges of the duct.  There are three main reasons for this: $i$) the flow is largely determined by flow details at the edges of the duct (e.g. the analytical solutions obtained using either two-layer hydraulics or the HGV-A approximation are highly dependent on the boundary conditions there); $ii$) the HGV-A approximation formally breaks down immediately outside the duct; $iii$) instabilities first emerge at the edges of the duct where the non-linear terms are larger since both $u$ and $\partial u /\partial x$ grow as $|x|\to1$ (this can be deduced from conservation of mass and the shape of the interface). Hence, further knowledge of the flow and the experimental setups at the edges of the duct is needed to clarify these points. Notice, for example, that \citet{Meyer2014} used two different types of edges in the duct, and any of these shapes are quite different than the solid long vertical walls in the simulations by \cite{Kaptein2020}. Furthermore, more information about the flow at the edges of the duct can help understand how it transitions from the HGV-A approximation to being hydraullically controlled.}

\CB{In spite of the differences between the setups, Figure \ref{fig:regimes} suggests that the parameter $\Fr^*$ does quantify the forcing for a given setup. Let's recall that this parameter groups together the control parameters of the setup: $\Rey_g$, $A$ and $\theta$ with the latter incorporated in the value of $K$. It was already known that $\Rey_g A^{-1}$ is the control parameter for horizontal ducts \citep{Hogg2001}. In this way, $\Fr^*$ is an equivalent number which is only re-scaled by dividing by $K$ to account for the relative inclination of the duct with respect to its internal angle (i.e. depending on $\theta/\alpha$). It would seem that $K$ is an intrinsic parameter of SID setups. A different way to analyze the problem is in terms of a generalized gravitational Reynolds number for inclined ducts. If the $\Rey_g$ quantifies the forcing due to the density differences, a generalized gravitational Reynolds number
\begin{equation}
    \Rey_{g,\theta}=  \dfrac{K(\theta=0)}{K}\Rey_g\approx \dfrac{131}{K} \Rey_g
\end{equation}
can be defined, in a similar way as done by \cite{Kaptein2021ReproducingDomains}. In this way, $\Rey_{g,\theta}$ incorporates both aspects of the forcing: the density difference between the tanks and the tilting of the duct.}

\CB{We have shown that the transition from laminar to turbulent in a (long) SID experiment with $\Sc\gg1$ depends solely of $\Fr^*$. For weak forcing (small $\Fr^*$ values) the flow is laminar, and it can be described by HGV-A approximation. As the forcing (the value of $\Fr^*$) is increased, the flow starts to develop non-linear and time-dependent characteristics (waves, thinner boundary layers, turbulence) that result in additional vertical viscous momentum diffusion that, for $\Fr^*>1$, compensates the increase in forcing. In this way, the value of $\Fr$ can remain constant in agreement with (frictional) two-layer hydraulic theory \citep{Gu2005}.} 


\section{Conclusions}

\CB{In the current paper, we present an analytical solution for the laminar flow in stratified inclined duct (SID) experiments within the HGV-A approximation, where the acronym stands for hydrostatic/gravitational/viscous balance in momentum and advective for density. This approximation is derived for long ducts ($A=L/H\gg 1$) and small-inclination angles $\theta$ (including $\theta=0$ and slightly negative angles). The Reynolds number has to be small enough for viscosity to dominate over inertia in the along-channel momentum equation, but diffusion is negligible in the density transport equation. This combination is only observed if $\Sc \gg 1$. Under these conditions, the flow gets organized in two homogeneous layers and the non-dimensional volume flow rate is given by $\Fr^*=\Reg (K\,A)^{-1}$, where $\Reg$ is the gravitational Reynolds number and $K$ is a geometrical parameter that depends exclusively on $A\sin \theta\approx \theta/\alpha$ with $\alpha$ the internal angle of the duct.}

\CB{A comparison with results from laboratory experiments has demonstrated that the analytical solution describes well the flow in the laminar regime and the time-averaged flow in the Holmboe waves regime. The HGV-A approximation presents the relevant balances in these two regimes, and it clearly exposes the two forcing mechanisms: a baroclinic pressure gradient due to the inclined interface and gravity in the along-duct direction due to the tilt of the duct. For $\theta/\alpha\gtrsim 1$, the inclination of the interface might be neglected, but for $\theta/\alpha\lesssim 1$, it is a crucial or even the main component of the forcing despite the fact the slope of the interface is small (e.g. $S'\approx 48/(131 A)$ for $\theta=0$). To make the link between horizontal and inclined ducts, \it is crucial to include both of these terms.}

\CB{Finally, the curves given by $\Fr^*=\text{const.}$ describe the regime transitions observed experimentally for horizontal and slightly (both positively and negatively) inclined ducts. As the value of $\Fr^*$ is increased, the forcing magnitude is increased. First, the HGV-A approximation breaks down as a critical $\Fr^*$-value is approached and reached. Successive transitions from laminar flow, to interfacial waves, to intermittent turbulence and sustained turbulence are needed to increase momentum diffusion to keep the internal Froude number from exceeding the maximum value imposed by maximum exchange theory. In this way, the curves $\Fr^*=\text{const.}$ provide a solution to the long-standing problem of finding the curves describing the regime transitions spanning horizontal and inclined ducts.}

\section*{Acknowledgements} 
This research was funded by STW now NWO/TTW (the Netherlands) through the project “Sustainable engineering of the Rhine region of freshwater influence” (no. 12682). The authors would like to acknowledge A. Lefauve and P.F. Linden for making the experimental data available, and for fruitful input and discussions. 

\section*{Declaration of interests}
The authors report no conflict of interest.
\appendix

\bibliographystyle{jfm}
\bibliography{exchange_flows.bib}

\end{document}